%% file: Main.tex
\documentclass[journal]{IEEEtran}
\usepackage{amsmath,amssymb,amsbsy,amsxtra,graphicx}
\usepackage{array,multirow}
\usepackage{booktabs}
\usepackage{url}
\usepackage{cite}
\usepackage{bm}
\usepackage{caption}
\usepackage{subfig}
\usepackage{epstopdf}
\usepackage{color}
\usepackage{balance}
\usepackage{graphicx}
\usepackage[table,xcdraw]{xcolor}
\usepackage{amsmath,amssymb}

\usepackage{float}
\usepackage[flushleft]{threeparttable}
\usepackage{dblfloatfix}
\graphicspath{{./Figures/}}
\usepackage{comment}
\usepackage{hyperref}
\usepackage{tabularx}

\usepackage{xcolor}
\usepackage{soul}

\begin{document}

\bstctlcite{IEEEexample:BSTcontrol}

\title{Image-to-Images Translation for Multi-Task Organ Segmentation and Bone Suppression in Chest X-Ray Radiography}

\author{

Mohammad Eslami,~\IEEEmembership{Member,~IEEE,}
        Solale Tabarestani,~\IEEEmembership{Student-Member,~IEEE,}
        Shadi Albarqouni,~\IEEEmembership{Member,~IEEE,}
        Ehsan Adeli,~\IEEEmembership{Member,~IEEE,}
        Nassir Navab,~\IEEEmembership{Member,~IEEE,}
        and~Malek Adjouadi,

\thanks{This work is supported by National Science Foundation (NSF) under NSF grants CNS- 1920182, CNS-1532061, CNS-1551221. }
\thanks{M. Eslami, S. Tabarestani and M. Adjouadi are with Center for Advanced Technology and Education (CATE), Florida International University, Miami, USA. This work is initiated while Eslami was a visiting scholar at CAMP-TUM (e-mails: meslami@fiu.edu, staba006@fiu.edu, adjouadi@fiu.edu).}
\thanks{S. Albarqouni and N. Navab are with Computer Aided Medical Procedures and Augmented Reality (CAMP), Technical University of Munich (TUM), Munich, Germany. (e-mails: shadi.albarqouni@tum.de, nassir.navab@tum.de)}
\thanks{E. Adeli is with Stanford  University,  Stanford,  CA, USA. (e-mail: eadeli@stanford.edu)}
\thanks{-}

}

\maketitle


\input{SecAbstract}

\IEEEpeerreviewmaketitle

\input{SecIntroduction}

\input{SecRelatedWorks}
\input{SecMethod}

\input{SecExperiments}

\input{SecConclusion}

\section*{Acknowledgments}
This work is supported by National Science Foundation (NSF) under NSF grants CNS- 1920182, CNS-1532061, CNS-1551221. 

\bibliographystyle{IEEEtran}

\end{document}

%% file: SecAbstract.tex
\begin{abstract}

Chest X-ray radiography is one of the earliest medical imaging technologies and remains one of the most widely-used for diagnosis, screening, and treatment follow up of diseases related to lungs and heart. The literature in this field of research reports many interesting studies dealing with the challenging tasks of bone suppression and organ segmentation but performed separately, limiting any learning that comes with the consolidation of parameters that could optimize both processes. This study, and for the first time, introduces a multitask deep learning model that generates simultaneously the bone-suppressed image and the organ-segmented image, enhancing the accuracy of tasks, minimizing the number of parameters needed by the model and optimizing the processing time, all by exploiting the interplay between the network parameters to benefit the performance of both tasks. The architectural design of this model, which relies on a conditional generative adversarial network, reveals the process on how the well-established \textit{pix2pix} network (\textit{image-to-image} network) is modified to fit the need for multitasking and extending it to the new \textit{image-to-imag\underline{es}} architecture. The developed source code of this multitask model is shared publicly on Github\footnote{\href{https://github.com/mohaEs/image-to-images-translation}{https://github.com/mohaEs/image-to-images-translation}} as the first attempt for providing the two-task \textit{pix2pix} extension, a supervised/paired/aligned/registered \textit{image-to-imag\underline{es}} translation which would be useful in many multitask applications. Dilated convolutions are also used to improve the results through a more effective receptive field assessment. The comparison with state-of-the-art algorithms along with ablation study and a demonstration video\footnote{\href{https://youtu.be/J8Uth26_7rQ}{https://youtu.be/J8Uth26\_7rQ}} are provided to evaluate efficacy and gauge the merits of the proposed approach.
\end{abstract}

\begin{IEEEkeywords}
chest X-Ray, CXR imaging, organ segmentation, bone suppression, multitask deep learning, image-to-image translation, image-to-images translation, pix2pix. 
\end{IEEEkeywords}

%% file: SecIntroduction.tex
\section{Introduction}\label{Introduction}
Chest radiography, also called chest X-ray or CXR, is one of the most affordable and widely used medical imaging modality, which has significant practical implications in the diagnosis and screening of the thorax region, the organs and bone structure within it. Over 2 billion procedures per year are performed using this technology for the purpose of medical diagnosis of a variety of  diseases, such as pneumonia, tuberculosis, lung cancer, and heart failure. Moreover, chest radiography remains the most prevalent screening test for pulmonary disorders \cite{general-1,general-nature,  general-2, pulmonary_abnormalities_1,pulmonary_abnormalities_3}. 
However, due to overlapping organs, low resolution and subtle anatomical shape and size variations, interpreting CXR images accurately remains challenging and requires a well-trained staff. On the other hand, managing a large number of CXR images each day results in high workloads for the radiography staff, yielding a tedious process fraught with setbacks and errors in diagnosis and in assessing adequately treatment follow up.  It is reported that almost 90 percent of mistakes in pulmonary tumor diagnosis could be associated with the CXR screening of images \cite{chest-512}. 
Therefore, many efforts have been devoted to the development of automated computer-based methods to improve accuracy in diagnosis and in finding any abnormalities that may otherwise be left undetected \cite{general-diamant2017chest, general-qin2018computer, general-tmi,pulmonary_abnormalities_2}.

There is considerable literature focusing on CXR image analysis. Among the more recent work on chest radiography, a team from Stanford \cite{general-Chest-Andrew} proposed a convolutional neural network called \textit{CheXNeXt} as a deep learning algorithm to concurrently detect the presence of 14 different pathologies such as pneumonia, fibrosis, emphysema, and nodules in frontal-view chest radiographs, among others. The \textit{CheXNeXt} algorithm achieved promising results in identifying abnormalities at a performance level that was comparable with the diagnostic accuracy of radiologist practitioners. 
Four different deep learning based methods are investigated in \cite{net_radiology} and compared with radiology experts.
In another study \cite{chest-512}, Gozes and Greespan proposed a method to improve the contrast of lung structures in CXR images leading to better accuracy in nodule(s) detection; while Wang and Chia proposed a deep neural network they named \textit{ChestNet} \cite{general-wang2018chestnet} for enhanced diagnosis of diseases on chest radiography. 
Moreover, an interesting multi-resolution convolutional network for chest X-ray radiograph lung nodule detection is proposed by Li et al. \cite{nodule_deep}.  


The aim of this work is to construct a multitask learning framework using deep learning techniques that address in an effective way the two challenging tasks of organ segmentation and bone suppression simultaneously. 
Organ segmentation is used for computer-aided detection and diagnosis while bone suppression enhances the visibility of the disease effects, e.g. nodules particularly on the lung region. Baltruscha et al. show that both of these tasks can improve the diagnosis rate by machine \cite{BSE-4}.
In order to incorporate the multitask objective, a new \textit{image-to-imag\underline{es}} translation machine is proposed based on the well-known \textit{pix2pix} network which is known for its promising results in the domain of \textit{image-to-image} translation and segmentation \cite{pix2pix}. 
For this reason, the \textit{pix2pix} network and its implementation are modified to fit the need for multitasking (\textit{pix2pix MT}). As far as the authors know, the proposed network is a first attempt at expanding the application of \textit{image-to-image} network to \textit{image-to-imag\underline{es}} with the ability to generate more than one desired output at once.
Furthermore, the dilated convolution technique \cite{dilation} is employed in specific layers of the generator, which is shown to improve further the results.
Hence, this design is referred to as the \textit{pix2pix MTdG} model given the implication of \textit{multitask pix2pix} and the inclusion of the dilation property in the generator.
More specifically, by feeding a CXR image to the \textit{pix2pix MTdG} network, the proposed model will generate automatically two output images simultaneously, which are the image of the bone suppressed lungs and the image containing the segmentation masks of the heart and lungs. 
Experimental results show that the \textit{MTdG} network yields promising and comparable results to the state-of-the-art methods that deal with these tasks individually. Results which are evaluated with several metrics and verified using 5-fold cross-validation along with the significance test exhibit promising outputs for both tasks. 
Moreover, the conceptual design of the model can be generalized to extend to other applications. To confirm this assertion, two different applications involving 1) neuroimage modality conversion for cross-modality generation of  T2-flair and T1-inverse from the T1 input image, and 2) low-dose Computed Tomography (LDCT) for image enhancement and segmentation of kidneys are addressed to showcase the merits of the proposed \textit{image-to-images} network in such critical applications where two desired outputs can be obtained simultaneously with improved accuracy and with better system efficiency.


The contributions of this work can be summarized as follows: 1) Design and implementation of a multitask network that for the first time augments the traditional \textit{image-to-image} translation model to an \textit{image-to-images} translation model while improving both accuracy and computational efficiency of the \textit{multitask pix2pix} model.
2) An architectural design of the model that allows for two critical tasks of CXR image analysis, namely bone suppression and organ segmentation, to be performed simultaneously through the use of efficient network parameters verified and augmented by an ablation study. 
3) A generalized construct making the model more amenable to other application domains, where the results of two more medical applications, brain MRI cross-modality generation and low-dose CT image enhancement and segmentation, are provided in support of this assertion. 
4) All the software code for the different variations of this work are publicly shared online including \textit{multitask pix2pix} for the research community to replicate such work or extend its research potential to other applications.

The rest of this paper is organized as follows: Section \ref{sec-related} reports the literature review of related work. Section \ref{sec-method} explains the proposed method, specifies the material used in conducting this study and the evaluation strategy that was used to assess its merits. The experimental results are presented and discussed in section \ref{sec-results} which include a section on method generalization and future work. Finally, the conclusion section \ref{sec-conclusion} provides a retrospective of what was accomplished through this proposed novel approach.

%% file: SecRelatedWorks.tex
\section{Related Work} 
\label{sec-related}

\subsection{Task 1: Organ Segmentation}

Organ segmentation is one of the most difficult tasks in medical imaging due in large part to the elusive thresholding process and the ubiquitous presence of noise \cite{malek-1,malek-2}, but remains an essential task for delineating the anatomical structures of organs and hence for detecting abnormalities such as enlarged heart or collapsed lungs. It should be noted that when performing segmentation in chest radiography, one would also need to contend with the different shape variations in organs due to age, gender, disease and other health-related issues. 

Mansoor et al. \cite{seg-mansoor2015segmentation} presented a comprehensive survey discussing the challenges and accomplishments of the different segmentation methods for lungs which are reported in the literature. By considering the CXR as the imaging modality, several deep learning models based on fully convolutional networks have also been investigated. For instance, a network called \textit{InvertedNet} is proposed to segment the heart, left and right clavicles, and lungs \cite{seg-InvertedNet}. 
The well-established \textit{U-Net} architecture has been utilized for segmenting the chest region yielding promising results \cite{seg-u-net-scia, seg-u-net-brazil, seg-u-net-islam2018towards}. 
A model called structure correcting adversarial network (\textit{SCAN}) was proposed as a generative adversarial network that uses convolutional layers for heart and lungs segmentation \cite{seg-SCAN}. Another method which incorporates two networks is proposed by \cite{seg_2019} with one network used for the initial segmentation process and the second for fine-tuning and correcting the initial results.
Moreover, traditional feature extraction methods are widely used for CXR imaging applications \cite{seg-mansoor2015segmentation}. In \cite{seg-Ibragimov}, Ibragimov et al. proposed an approach for lung segmentation and landmark detection based on Haar-like features, a random forest classifier, and spatial relationships among landmarks. A hierarchical lung field segmentation based on the joint shape and appearance sparse learning is proposed in \cite{seg_sparse}, and an atlas-based method is presented in \cite{seg_atlas}.

\subsection{Task 2: Bone and Rib Suppression}
In chest X-ray images, the bone structure in the chest area is usually visible, which makes it hard for a radiologist to examine thoroughly the organs and assess any effects of a given disease accurately. Organs' visibility is effective for pulmonary abnormalities screening and detection \cite{pulmonary_abnormalities_3}. Consequently, bone and rib suppression is an essential pre-processing step in order to suppress the appearance of bones in the chest X-ray images.
One way to tackle the aforementioned problem is to utilize dual-energy subtraction (DES) imaging \cite{DES}. The DES imaging technique captures two or three radiography scans with two or three different energy level of X-ray exposures. 
The captured images either highlight the soft tissues or bones based on the energy levels. Thus, the suppressed bone image will be estimated by combining the acquired images which include both the soft tissue-selective images and the bone-selective images \cite{DES-2}.  Although effective in delineating the bone structure in the chest area, the DES imaging process has a number of shortcomings, among them is its more invasive nature due to the higher radiation dose and the presence of artifacts introduced in the acquisition process due to the effect of heartbeats.

Because of these aforementioned reasons, suppressing the bones in CXR images via traditional image processing techniques is considered safer and is shown to be more effective at overcoming the main challenges faced in CXR images. 
Along this line of research, a cascaded convolutional neural networks architecture (called \textit{CamsNet}) \cite{BSE-2} is proposed to predict the bone gradients in CXR images progressively with the ability of suppressing the bones as a consequence of these determined gradients. Convolutional neural filters are exploited in \cite{bse_CT} and are shown to be effective for bone suppression as well. 
Another recent method is developed by Chen \textit{et al.} \cite{bse-tmi} which anatomically compensates for the ribs and clavicles by specific multiple massive-training artificial neural networks (\textit{MTANNs}) combined with total variation (TV) minimization smoothing along with a post-processing by histogram-matching. 
In another study, Gusarev et al. proposed two deep learning architectures that perform bone suppression and create a soft tissue image. Considering bones as a noise level that is affecting these chest images \cite{BSE-3}, they tried to minimize the presence of this noise (i.e., bone) while still preserving the sharpness of the image for the eventual organ segmentation. In \cite{malek-1}, many of the noise suppressing methods reviewed shared the objective of removing as much of the noise as possible while preserving most of the relevant details in the image. 
Another bone suppression method, based on deep adversarial networks and 2D Haar wavelet decomposition, has been proposed in \cite{BSE-1}. Their method was mostly based on the theory of \textit{pix2pix} network \cite{pix2pix}, a well-known conditional generative adversarial network. The \textit{pix2pix} network is also used as the cornerstone of our proposed multitask model, which will be described in section \ref{sec-method}. 
Bone suppression is also used as a pre-processing step, where bone suppressed CXR images are then feed as input images to algorithms such as \textit{CheXNet} in order to enhance the segmentation process and improve as a consequence the results of the machine (automated) diagnosis \cite{bse-jsrt-usedin-BSE}.
The impact of bone suppression on machine diagnosis using deep learning networks have been thoroughly investigated and detailed in \cite{BSE-4} and \cite{bse-jsrt-usedin-deep}. There are also some commercially  available computer-aided detection (CAD)  systems such as \textit{Phillips} \cite{commercial_3}, \textit{ClearRead} \cite{commercial_1} and \textit{Caresteam} \cite{commercial_2}.

\subsection{Joint Tasks via Multitask Learning}

In multitask learning, multiple tasks are solved at the same time by exploiting commonalities and differences across tasks. In comparison to training separate single task models, the multitask scheme can result in the following improvements \cite{MTL-Survey}: 
1) Improvement in results: Most often, coupling tasks makes the overall system achieve better results with respect to the desired accuracy. For example, in \cite{HyperFace}, a multitask learning approach based on deep convolutional networks is proposed for facial landmark detection  with the auxiliary tasks of head pose estimation, gender classification and facial visibility, yielding more accurate results for each of the tasks. 
2) Improvement in learning efficiency: Efficiency of learning could include other important implementation aspects such as the number of required parameters, memory or storage requirements, computational time and training convergence rate. Obviously, fewer but optimal parameters and lower memory requirements are desirable in deploying such  algorithms on conventional devices such as mobile phones and PCs \cite{hardware}. 

Aside from the potential improvement in learning efficiency in the interplay between tasks, and the need for only half the weights required of the multitask model in contrast to the two tasks run separately, other benefits acquired through the intrinsic functions of the  multitasking model do not have straightforward reasons. For instance, an additional function can be used to act as a regularization mechanism in other machine learning problems and push the algorithm to find a solution on a smaller area of representations at the intersection of all tasks. Also, the feature selection and filter values can be reassessed and made more sufficient for addressing the nature of the inputs to the model. The overall motivation here is to be able to perform the two tasks of bone suppression and organ segmentation jointly via one deep network with the ability for improving both system efficiency and accuracy in the results.

\subsection{Image-to-Image translation: Applications and methods}

In general, the \textit{image-to-image} translation (I2I) is the process of translating an input image $X$ to a corresponding output image $Y$, and this correspondence could mean different things for the different context of the application at hand. Such I2I techniques could involve translations such as low-resolution $\Leftrightarrow$ high resolution, blurry $\Leftrightarrow$ sharp, thermal or grayscale $\Leftrightarrow$ color, synthetic $\Leftrightarrow$ real, low-dose rate (LDR) $\Leftrightarrow$ high-dose rate (HDR), noisy $\Leftrightarrow$ clean, image $\Leftrightarrow$ painting, day $\Leftrightarrow$ night, summer $\Leftrightarrow$ winter, bad weather $\Leftrightarrow$ good, foggy $\Leftrightarrow$ clear, semantic labeling (segmentation) $\Leftrightarrow$ realistic photo, aerial $\Leftrightarrow$ map, edges and sketch $\Leftrightarrow$ photo and so on, in which symbol $\Leftrightarrow$ shows the bidirectionality between desired task and context \cite{pix2pix, I2I-review-med}. I2I techniques are also used in medical imaging for segmentation, denoising, super-resolution, modality conversion, CT and MRI reconstruction, among others \cite{I2I-review-med}.

I2I has been studied for decades, and different approaches are reported on the basis of filtering, optimization, dictionary learning, deep learning, and more recently generative adversarial network (GAN). While deep learning methods omit the hand-crafted features and GAN methods omit the hand-crafted objective functions, they both remain the most promising methods in data science. GAN-based I2I research in computer vision has yielded different learning models, with a myriad of applications and promising outcomes. In general, there are two categories of methods and applications based on the relation between input and output images: \textit{1)  Unsupervised/Unpaired/Unaligned/Unregistered} such as style changing, photo to painting, hair/face and color-changing, weather changing, and \textit{2) Supervised/Paired/Aligned/Registered} such as supervised segmentation and labeling, denoising and super-resolution.   


The unpaired category is not relevant in the proposed application of CXR image analysis because the problem at hand is supervised with paired and aligned input/output images. Until now, \textit{pix2pix}\cite{pix2pix}, \textit{CRN}\cite{I2I-CRN}, \textit{BicycleGAN}\cite{I2I-BicycleGAN}, \textit{SIMS}\cite{I2I-SIMS}, \textit{SPADE}\cite{I2I-SPADE} and \textit{pix2pixHD}\cite{I2I-pix2pixHD} remain the most important methods for the paired category.  While \textit{pix2pix} and \textit{BicycleGAN} are dealing with a family of applications, others are just considering semantic labels to realistic photo translation. \textit{BicycleGAN} is an \textit{image-to-image} translation with potentially multiple outputs. For example, \textit{BicycleGAN} is able to analyze and translate a given night image to synthesized day images with different types of lighting, sky, and clouds. Each different possibility is generated by feeding a random noise sampled from a known distribution (e.g., a standard normal distribution) along with the input image. Therefore, \textit{pix2pix} is the only general method relevant to our problem, with bone suppression and organ segmentation being a paired/supervised/aligned/registered problem with no randomized output possibilities. 

%% file: SecMethod.tex
\section{Materials \& Methods} 
\label{sec-method}

\begin{figure*}[]
\captionsetup{justification=centering}
\minipage{0.49\textwidth}
  \includegraphics[width=\linewidth, trim={4.5cm 4.cm 5.5cm 3.5cm},clip]{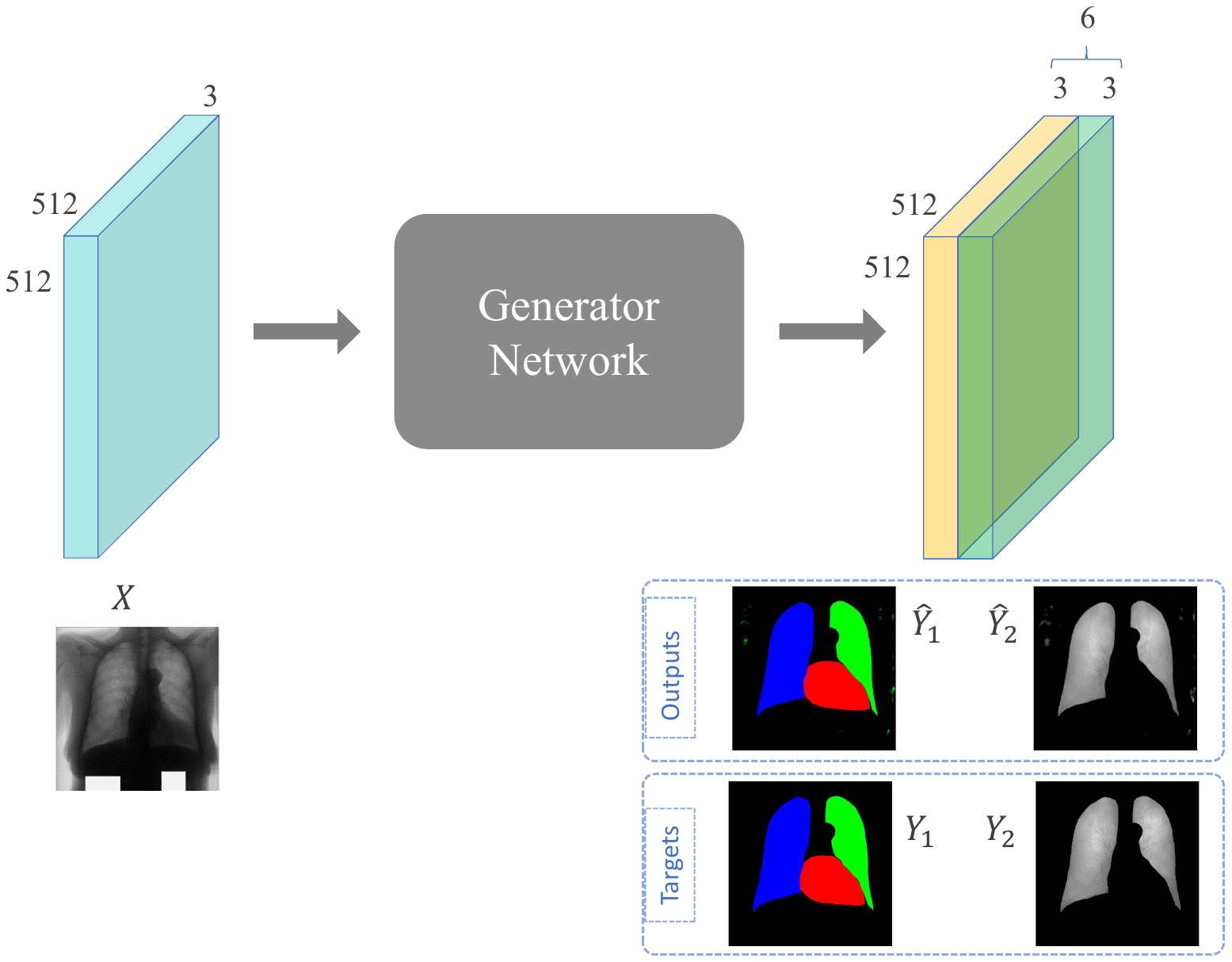}
  \caption*{a) Generator.}
\endminipage\hfill
\minipage{0.49\textwidth}
  \includegraphics[width=\linewidth, trim={4cm 3.5cm 5cm 3cm},clip]{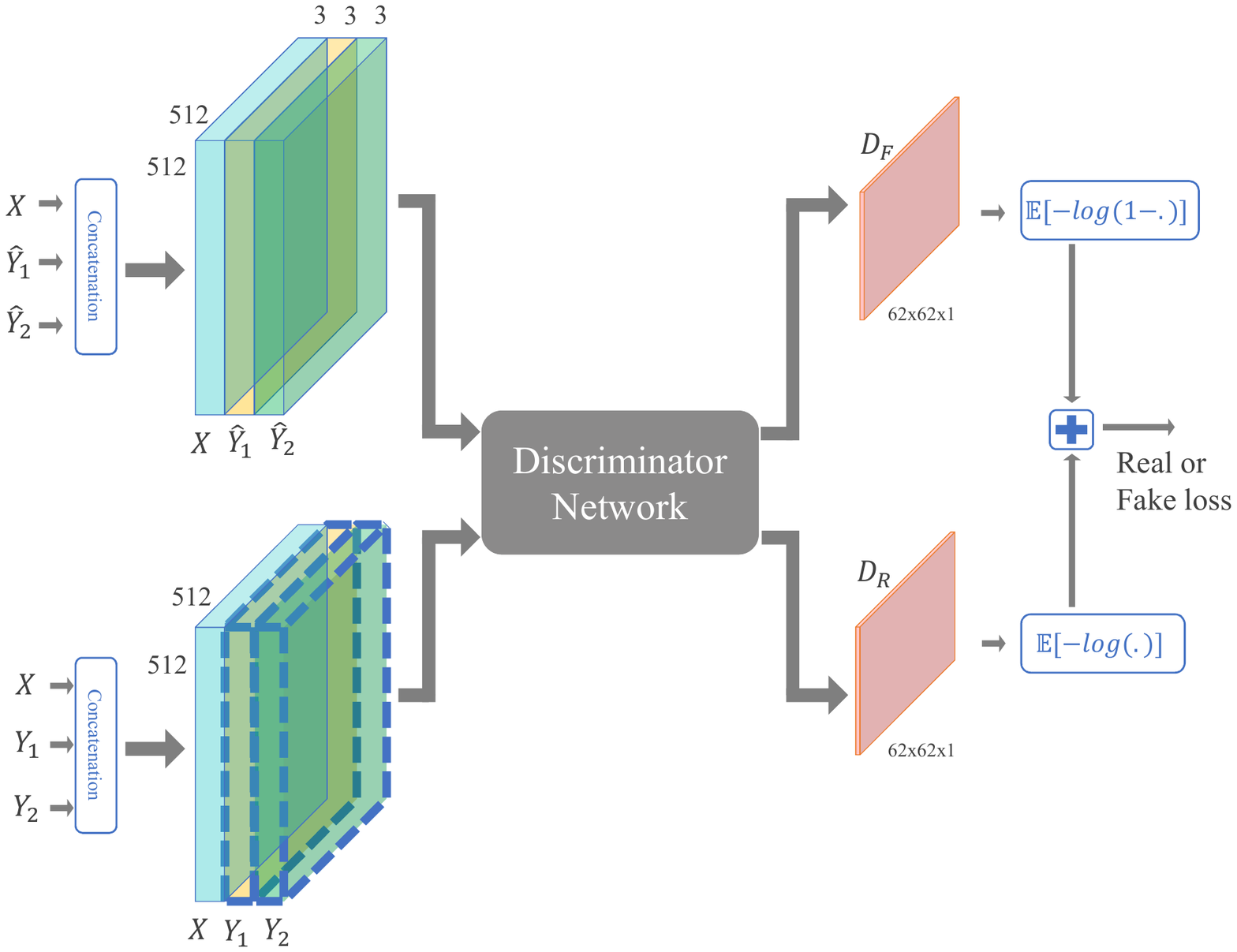}
  \caption*{b) Discriminator.}
\endminipage\hfill
  \caption{Architecture of the presented image to images translation, \textit{multitask pix2pix}. In this figure, $X, Y_1, Y_2,$ $\hat{Y}_1, \hat{Y}_1$ are the images of input CXR, targets of task 1 and task 2, output for task 1 and task 2, respectively. Notice that, all the images, input, output and target have three channels.}
  \label{fig:Model}
\end{figure*}

\subsection{Methodology}
\subsubsection{Background}

 In recent years, generative adversarial networks (GANs) and conditional generative adversarial networks (cGANs) have gained a lot of attention because of their superior performance in generation, segmentation, and translation empowered by an adversarial scheme \cite{cGAN}. The GAN architecture consists of two ‘adversarial’ models trained to work against each other: the generator aiming at generating an output and deceiving the discriminator and a discriminator component aiming at segregating the real output from the fake ones. In conditional mode (cGAN), both generator and discriminator are conditioned on ground truth labels or images. For example, in this study, the segmented organs and the bone suppressed images are the conditions and the generator is set up to generate this type of images. 



Generators of GANs are intended to learn the mapping from a random noise vector $z$ to an output image $y$, \textit{i.e.,} $G: z \Rightarrow y$ while cGANs are conditioned by an observed image $x$ \textit{i.e.,} $G: \ \{x,z\} \Rightarrow y$. The generator $G$ would learn to produce outputs, which could not be distinguished as “fake” images by an adversarially trained discriminator, $D$.
The objective of a GAN and of a conditional GAN can be expressed through equations \eqref{eq:Gan} and \eqref{eq:cGan} respectively, where $\mathbb{E}$ is the \textit{Expectation} over the population. 
Generator $G$ tries to minimize an objective function against an adversarial
$D$ which tries to maximize it, \textit{i.e.,} a \textit{minimax} game as $\hat{G}=arg \ min_G \ max_D \ \mathcal{L}_{GAN}$ and similarly $\hat{G}=arg \ min_G \ max_D \ \mathcal{L}_{cGAN}$. 
 
\begin{gather}
    \mathcal{L}_{GAN}(G,D) = \hspace{150pt} \nonumber \\ 
    \mathbb{E} \ [log D(x,y)]
    +
    \mathbb{E}  \ [log(1-D(G(x,z)))] \ \ \
    \label{eq:Gan} \\
    \mathcal{L}_{cGAN}(G,D) = \hspace{150pt}  \nonumber\\ 
    \mathbb{E} \ [log D(x,y)]
    +
    \mathbb{E}  \ [log(1-D(x,G(x,z)))]
    \label{eq:cGan}
\end{gather}

\subsubsection{Proposed Model}

A conditional generative adversarial network, called \textit{pix2pix}, is selected as an \textit{image-to-image} translation network to use and modify to meet the intended objectives of bone suppression and image segmentation \cite{pix2pix}. The \textit{pix2pix} is used in the proposed method, because 1) It is the main general \textit{image-to-image} translation method; 2) It shows promising prospects for accurate organ segmentation \cite{pix2pix,GanSeg2,GanSeg6,GanSeg9}; 3) It is intrinsically a collection of filters and would be reasonable to perform bone and rib suppression as an \textit{image-to-image} translation task \cite{BSE-1}.

The generator of \textit{pix2pix} contains an auto-encoding network of convolutional layers with skip connections. The discriminator is also a convolutional neural network (CNN) called \textit{PatchGAN} discriminator as introduced in \cite{pix2pix}, which attempts to determine whether each patch with size $n \times n$ in an image is real or fake, where $n$ can be much smaller than the full size of the image. Specifically, \textit{PatchGAN} discriminator is a CNN which produces a matrix of size $k \times k \times 1$ from an input tensor (or image) of size $N \times N \times *$ where $k=N/n$ and each element in the output matrix indicates the status of the corresponding receptive field on the input tensor (\textit{i.e.} a $k \times k$ \textit{PatchGAN} classifies $k \times k$ patches of the input image/tensor as real or fake). The input tensor for the \textit{PatchGAN} discriminator is a tensor built by concatenation of the input-target pair and the input-output pair for the discriminator to produce an estimation on how realistic they look \cite{pix2pix}.

Figure \ref{fig:Model} shows our model aiming to convert the input CXR image ($X$) into the desired output ($Y$), which is the concatenation of desired targets, $Y_1$ as the organs' segmentation masks and $Y_2$ as the bone suppressed CXR image, \textit{i.e.,} $Y: Y_1 \parallel Y_2$ where $\parallel$ shows concatenation in the channel axis.
The input and output tensors of the generator network are $X$ and $\hat{Y}: \hat{Y}_1 \parallel \hat{Y}_2$ where $\hat{Y}_2$ is the output image corresponding to the bone suppression task and $\hat{Y}_1$ is the output image corresponding to the organ segmentation task which include the masks for the heart (colored red), left lung (colored blue), right lung (colored green) and background (colored black). In fact, the generator creates a tensor with 6 channels, which are the concatenation of $\hat{Y}_1$ and $\hat{Y}_2$. 

The discriminator network acts in a similar fashion to \textit{PatchGAN} in order to produce two output matrices, $D_R$ and $D_F$ corresponding to the real and fake input tensors ($R$ and $F$). The fake input tensor ($F$) is a concatenation of CXR input image and outputs ($F: X \parallel  \hat{Y}_1 \parallel \hat{Y}_2$), and the real input tensor ($R$) is the concatenation of the CXR input image and targets ($R: X \parallel Y_1 \parallel Y_2$). If the discriminator is trained perfectly, it will create $D_R$ matrix of 1 values and $D_F$ matrix of 0 values. On the other hand, if the generator is successful in fooling the discriminator, $D_F$ would be a matrix of 1 values. The loss functions for training the generator and discriminator are as expressed in equations \eqref{eq:OurLoss-G} and \eqref{eq:OurLoss-D} where $\mid \ \mid_1$ defines the $L1$ distance or norm. In the training phase of the networks, for each batch feeding step: 1) The generator generates output images, 2) The discriminator looks at the real pair tensor ($R$) and the fake pair tensor ($F$) and produces an estimate on how realistic they look ($D_R$ and $D_F$), 3) The weights of the discriminator are then adjusted based on the $\mathcal{L}_D$, and 4) The generator's weights are then adjusted based on $\mathcal{L}_G$.

\begin{gather}
\mathcal{L}_G = \mathbb{E} [-log(D_F +\epsilon )] + \lambda \ \mathbb{E} [\mid Y - \hat{Y} \mid_1 ] 
\label{eq:OurLoss-G} 
\\ 
\mathcal{L}_D = \mathbb{E} [-( \ log(D_R + \epsilon) + log(1-D_F + \epsilon ) \ )]  
\label{eq:OurLoss-D}
\end{gather}

Furthermore, in order to to produce more efficient receptive fields, dilated convolutions \cite{dilation} are utilized in some specific layers of the generator. The encoder of the generator consists of 8 layers and dilated convolutions are used with dilation rate 2 in layers 2 through 7 in the proposed structure of the \textit{MTdG} network. The effects of using and not using the dilated convolutional layers are contrasted in the results section.

\subsection{Data}

The \textit{Japanese Society of Radiological Technology (JSRT)} is the only publicly available database where both desired tasks are available and hence most suitable for training and evaluating the proposed model. This dataset consists of CXR images collected by JSRT \cite{jsrt} and is publicly available in \cite{jsrt-data}. Segmentation masks for lungs and the heart were created later by \cite{jsrt-segments} and are now available in \cite{jsrt-segments-data}. The JSRT dataset comprises 247 CXRs, including images with and without lung nodules. All images have a resolution of $2048 \times 2048$ in gray scale with a color depth of 12 bits. 
While there is no publicly available dataset for bone and rib suppression based on DES, Juhász et al. developed a method for bone suppression \cite{bse-jsrt-alg}. Their results on the JSRT dataset have become publicly available in \cite{bse-jsrt-data}. This dataset is used to accomplish the second task of bone suppression as well. 

As noted in CheXNeXt and through other investigations, the $512\times512$ resolution is sufficient for classifying lung-related diseases and for localizing nodule(s) \cite{general-Chest-Andrew, chest-512, chest-512-2, chest-512-3}. This resizing of images helped to significantly minimize the computational requirements and, as the results will prove, high accuracy is maintained. Moreover, we anticipate that the resizing of images  to $512\times 512$ pixels could help in their effective use towards the development of new pre-processing methods for improving computer-aided diagnosis.
The image intensities are set up with an 8-bit grayscale resolution in the range from 0 to 255. 
In order to train the machine learning and especially the deep learning networks, it is essential to have enough number of samples that cover the different variations \cite{Augmentation1, Augmentation2}. Therefore, the original images along with their corresponding masks and suppressed bone images were augmented by rotating them via 10 and -5 degrees, along with translations of (30, 10) and (-20,-10) pixels in reference to the (x,y) coordinates.
Through this process, the size of the dataset has been increased by 5 times,  to a total number of 1,235 images along with their corresponding ground-truths for the two tasks. 


Furthermore, in order to assess the effectiveness of multitask \textit{image-to-images} translation in terms of its generalization to other applications, two additional experiments are included in the \textit{Generalization and Future Work} section. The dataset used for the low-dose CT (LDCT) experiment is from \textit{"Multi-Atlas labeling beyond the cranial vault"} challenge containing CT scans and corresponding segmentation labels of 13 abdominal organs of 50 subjects \cite{LDCT_data}. In order to simulate the LDCT scans from CT scans, the method based on additive Poisson noise on sonograms of CT scans is used \cite{LDCT_simulation}.  
The dataset for the neuroimaging experiment is \textit{MRBrainS18} challenge dataset containing multi-modal MRI brain images (T1, T1-inverse and T2-flair) with segmentation labels of gray matter, white matter, cerebrospinal fluid, and other structures on 3T MRI scans of the brain of seven subjects \cite{MRBrainS18}. Augmentation with the same configuration is also exploited for the LDCT and neuroimaging experiments.

\subsection{Implementation \& Evaluation}

The proposed model has been implemented and modified to comply with the multitasking scenario based on the publicly available \textit{pix2pix} code \cite{pix2pixCode}.
For validation purposes, both the extended code that supports all these different variations and the video showing this process at work are made available through the Internet to the research community. 
The intensity channel of the input CXR image is replicated to support the CNN 3-channel RGB input data expectations.
The size of the input/output images and kernel (or filter) are $512\times512\times3$ and $4\times4$, respectively and 5-fold cross-validation on subjects are considered as default. To make a fair comparison with the results of the current state-of-art-techniques, the resolution of  $256\times256$ and 3-fold cross-validation has been considered in this study as well.
The network has been implemented using Python and the Tensorflow library. All computations for training the network have been performed on a system equipped by NVIDIA GPU Quadro M6000 with 24 GB memory.
The parameter $\lambda$ in equation \ref{eq:OurLoss-G} is set to 10 and the learning rate of the Adam optimizer is set to 0.0002. The training is stopped when the L1 loss reaches almost $0.005$.
Because of a limited number of subjects, leaving-one-subject-out cross-validation scheme is used for the LDCT and neuroimaging experiments. 
For comparative assessment of the results obtained with the different methods, the average, standard deviation, box plots including median, percentiles, and outliers along with t-test for statistical significance are considered regarding the different metrics.

%% file: SecExperiments.tex
\section{Results and discussion}
\label{sec-results}
In this section, the experimental results for both tasks are reported and compared with state-of-the-art methods in the first two subsections. This is followed by a subsection which provides the performance and qualitative analysis, investigates the ablation study and demonstrates how this model is amenable to other application domains (i.e. generalization of the model) with preliminary results obtained on low-dose CT image processing and on a neuroimage translation problem. The generalization subsection shows the benefits of multitasking in comparison with single tasking and the potential for future work of \textit{pix2pix MTdG} towards resolving other \textit{image-to-imag\underline{es}} translation problems. Notice that, for simplification in plotting the figures \textit{p2p} notation is used instead of \textit{pix2pix}.   


\subsection{Task 1: Organ Segmentation}

\begin{figure}[!t]
\centering
\captionsetup{justification=centering}
\minipage{0.4\textwidth}
  \includegraphics[width=\linewidth]{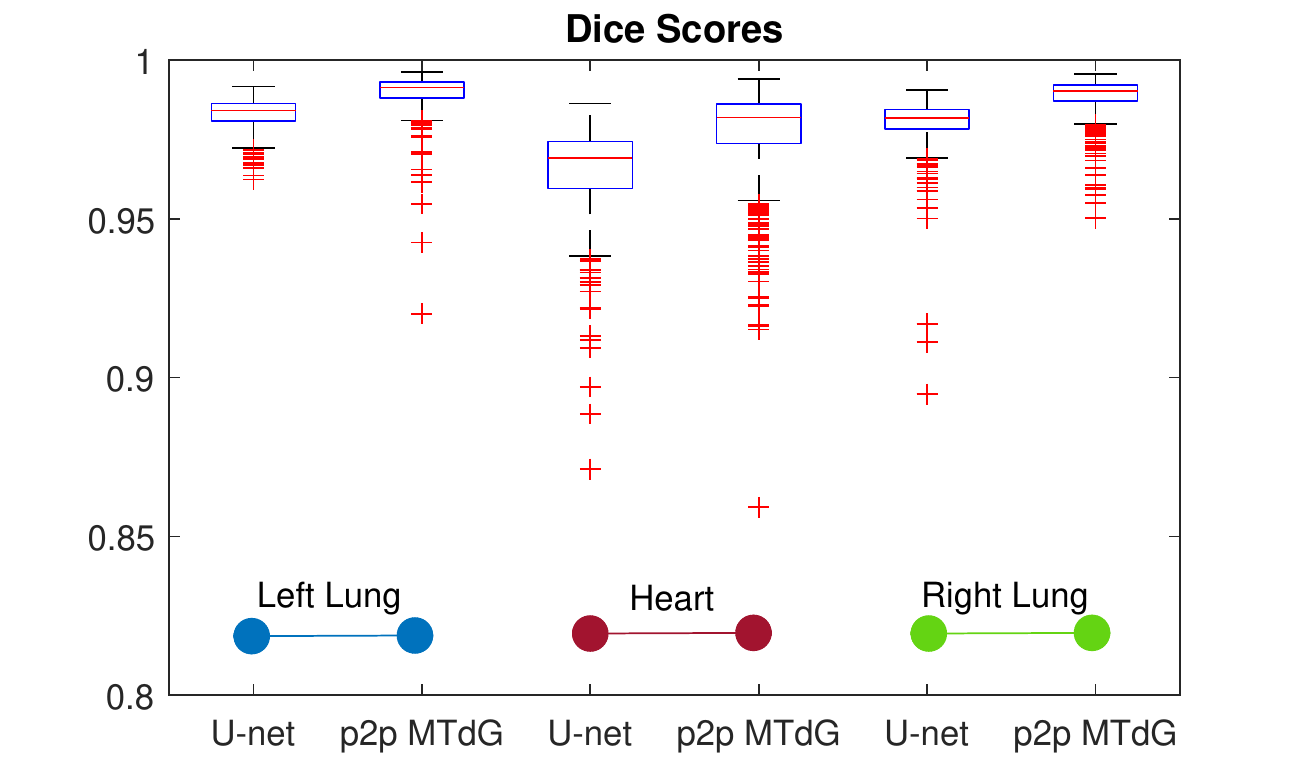}
\endminipage\vfill
\minipage{0.4\textwidth}
  \includegraphics[width=\linewidth]{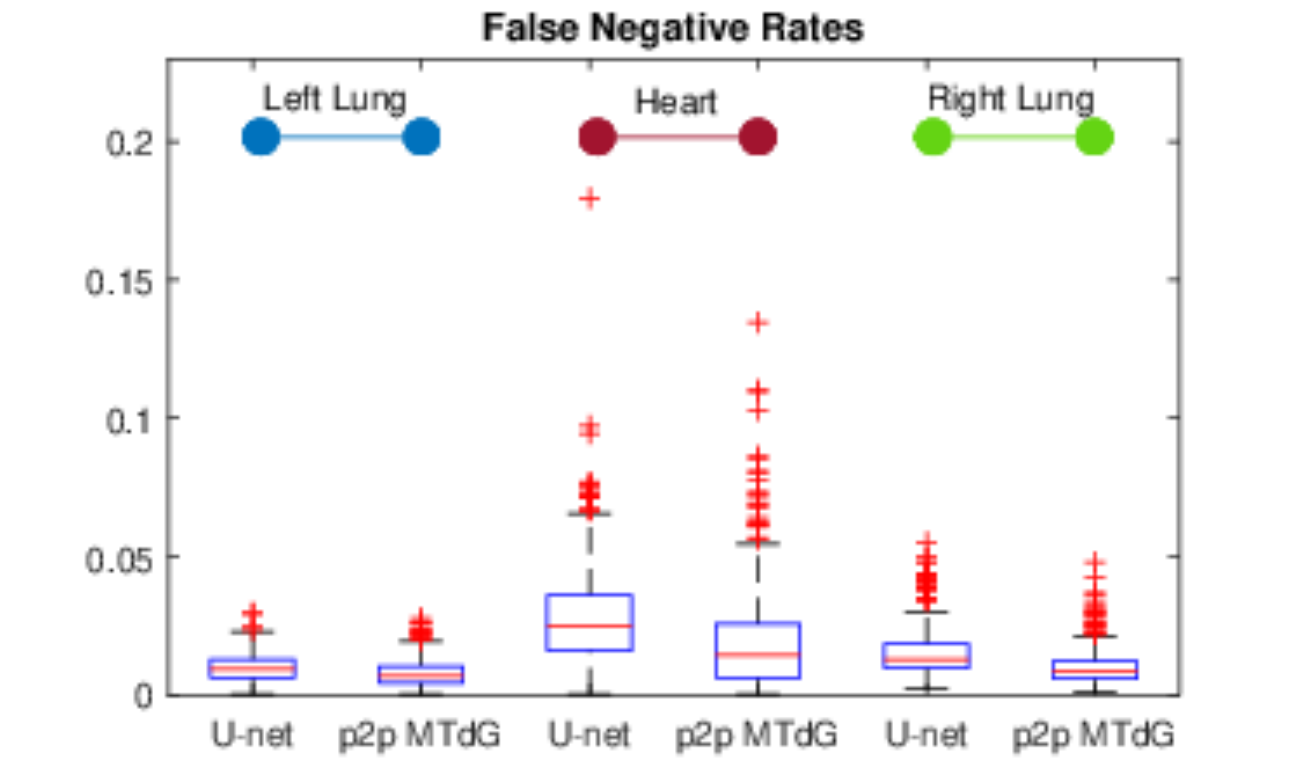}
\endminipage
  \caption{The box plot of the segmentation accuracy achieved by \textit{u-net} and  \textit{pix2pix MTdG}. \\ up) Dice Scores, Down) False Negative Rates.}
  \label{fig:compare-T1}
\end{figure}

The segmented regions of the heart, the left lung and right lung generated by the model as output masks, are associated with the same regions in the ground-truth by using standard evaluation metrics in image processing, namely the Dice and Jaccard scores, false-negative rate (FNR) and false-positive rate (FPR). 
The Jaccard index is a metric that measures the percent overlap between the target ground-truth mask (GT) and our prediction mask (PM). 
The overlap area between GT and PM would be the true positive area (TP). The area which is predicted in PM but is not in PM is a false-positive (FP) and inversely for false-negative (FN) defining the area which is in GT but not in PM.  
Jaccard metric is closely related to the Dice coefficient, which is not as easily described geometrically. 
The false-positive rate indicates the area ratio of predicted mask which had no associated ground truth mask, and similarly false-negative indicates the area ratio of the ground truth mask which had no associated predicted mask. 
These standard evaluations metrics can be expressed mathematically as follow where symbol ! defines the binary negation operator: 

\begin{gather}
Dice=\frac{2 \times (PM \cap GT)}{PM + GT}= \frac{2 \times TP}{TP + FP + TP + FN} \\
Jaccard=\frac{PM \cap GT} {PM \cup GT} = \frac{TP}{TP + FP+ FN}  
\\ 
FNR=\frac{!PM \cap GT}{GT}= \frac{FN}{TP + FN} 
\\ 
FPR= \frac{PM \cap !GT}{GT} = \frac{FP}{TP + FN}
\label{eq:SegScores}
\end{gather}

\begin{figure}[]
\centering
\captionsetup{justification=centering}
\minipage{0.5\textwidth}
  \includegraphics[width=1\linewidth]{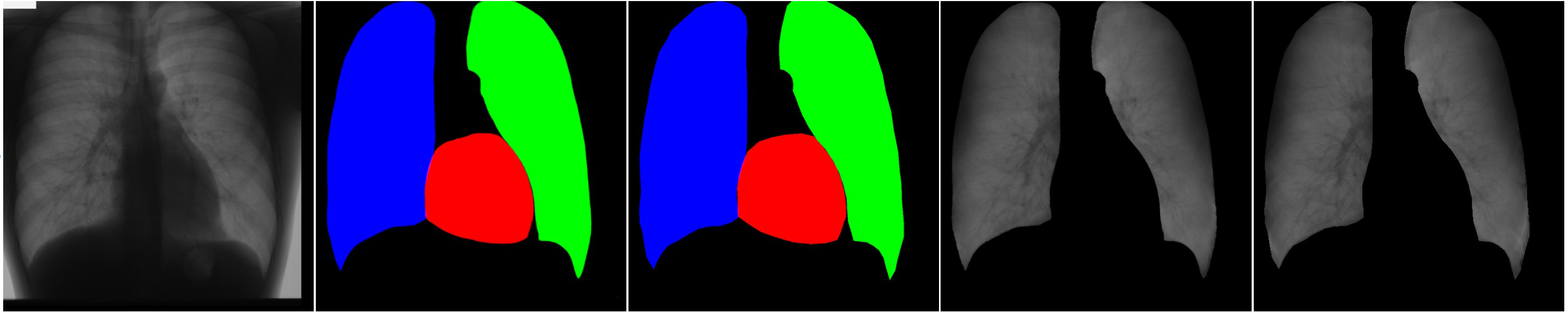}
\endminipage\vfill
\minipage{0.5\textwidth}
  \includegraphics[width=1\linewidth]{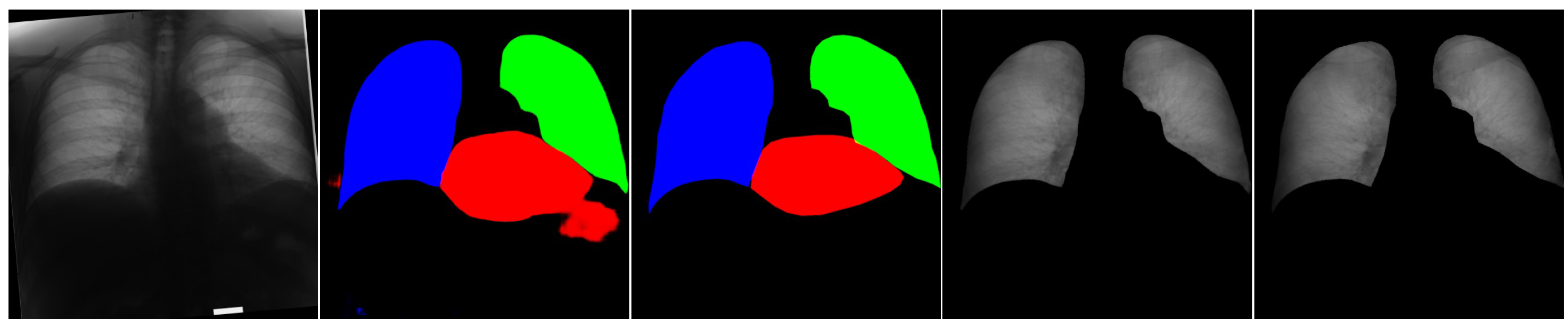}
\endminipage\vfill
  \caption[Caption for LOF]{Results by proposed method \textit{pix2pix MTdG} regarding to the best (top) and worst  (bottom) Dice scores. \\
  \footnotesize{Columns left to right are, input image, segmentation result, segmentation target, bone suppression result, and bone suppression target. \\ 
  Top) Best Dice: average of Dice, Jaccard, FPR and FNR are 0.99, 0.99, 0.01 and 0.01. \\ 
  Bottom) Worst Dice: average of Dice, Jaccard, FPR and FNR are 0.94, 0.90, 0.12 and 0.01. } \\ 
  }
  \label{fig:image-dice}
\end{figure}

The well-known U-net method \cite{u-net} is used as the competitor due to its state-of-the-art performance in various segmentation applications \cite{u-net-nature} as well as for the reported CXR segmentation studies with promising results \cite{seg-u-net-brazil, seg-u-net-scia, seg-u-net-islam2018towards}.
The average and standard deviation of the segmentation results, by all metrics, are summarized in Table \ref{tab:segmentation}. 
The best achieved results, highlighted in blue, demonstrate that the \textit{multitask pix2pix} with an embedded dilation in the generator (\textit{MTdG}) surpasses the \textit{u-net} method.
For further statistical investigation, figure \ref{fig:compare-T1} shows the box plots of the segmentation scores evaluated using the Dice and false-negative rate of the heart, left and right lung for the \textit{u-net} and \textit{pix2pix MTdG}. 
The inner line, the bottom and top edges of the box indicate median, the 25th, and 75th percentiles, respectively. While the whiskers are extended to the most extreme not outlier data points, while the outliers are plotted individually using the '+' symbol. 
As an example, the segmentation result of the \textit{pix2pix MTdG} for the best and worst achieved Dice scores are shown in Figure \ref{fig:image-dice}.

\begin{table}[]
\centering
\captionsetup{justification=centering}
\minipage{0.5\textwidth}
\centering
\caption{Segmentation results of different methods while the best scores are colored blue. }
\label{tab:segmentation}
\begin{tabular}{|c|c|c|c|c|c|c|}
\hline
                                  &                   & \textbf{u-net}            & \textbf{pix2pix MTdG}  & \textbf{p-value} \\ \hline 
\multirow{4}{*}{\textbf{Dice}}    & Left lung         &    0.983 $\pm$ 0.005          &          0.990 $\pm$ 0.006   &        1.2 e-93    \\ 
                                  & Heart             &      0.965 $\pm$ 0.014        &       0.977 $\pm$ 0.015          &     3.6 e-57  \\ 
                                  & Right lung        &        0.980 $\pm$ 0.009      &              0.988 $\pm$ 0.006     &  5.5 e-99   \\ 
                                  & \textbf{Average}  &      0.976 $\pm$ 0.007        &   \textcolor{blue}{  0.985 $\pm$ 0.007 }    &    -    \\ \hline
\multirow{4}{*}{\textbf{Jaccard}} & Left lung         &        0.967 $\pm$ 0.010      &          0.980 $\pm$ 0.011             &  7.7 e-97 \\ 
                                  & Heart             &        0.933 $\pm$ 0.026      &           0.956 $\pm$ 0.027      &   3.8 e-59    \\ 
                                  & Right lung        &     0.961 $\pm$ 0.017         &     0.977 $\pm$ 0.012             &   6.9 e-105  \\ 
                                  & \textbf{Average}  &       0.953 $\pm$ 0.013       &   \textcolor{blue}{ 0.971 $\pm$ 0.013 }       &      -     \\ \hline
\multirow{4}{*}{\textbf{FNR}}     & Left lung         &       0.010 $\pm$ 0.005       &        0.008 $\pm$ 0.005               &  8.8 e-18 \\ 
                                  & Heart             &       0.028 $\pm$ 0.017       &        0.018 $\pm$ 0.017     &      7.1 e-24     \\ 
                                  & Right lung        &        0.015 $\pm$ 0.008      &          0.010 $\pm$ 0.006        &     3.0 e-45 \\ 
                                  & \textbf{Average}  &      0.017 $\pm$ 0.006        &        \textcolor{blue}{ 0.012 $\pm$ 0.007  }        &  -    \\ \hline
\multirow{4}{*}{\textbf{FPR}}     & Left lung         &          0.024 $\pm$ 0.013    &        0.013 $\pm$ 0.013               &  3.0 e-73\\ 
                                  & Heart             &       0.043 $\pm$ 0.033       &         0.028 $\pm$ 0.029        &   2.3 e-22   \\ 
                                  & Right lung        &      0.026 $\pm$ 0.017        &         0.013 $\pm$ 0.012      &       8.3 e-67 \\ 
                                  & \textbf{Average}  &     0.031 $\pm$ 0.015         &            \textcolor{blue}{0.018 $\pm$ 0.013}     &    -   \\ \hline
\end{tabular}
\endminipage
\end{table}

\begin{table*}[]
\minipage{0.9\textwidth}
\begin{center}
\begin{threeparttable}
\caption{Comparison the segmentation results of different methods on JSRT dataset.}
\label{tab:seg-comp}
\begin{tabular}{|c|c|c|c|c|c|c|c|}
\hline
\textbf{Method}                     & \textbf{Image Size} & \textbf{Augmentation} & \textbf{Evaluation scheme} & \textbf{Dice} & \textbf{Jaccard} & \textbf{Dice} & \textbf{Jaccard} \\ \hline
                                    &                     &                       &                     & \multicolumn{2}{c|}{Lungs}       & \multicolumn{2}{c|}{Heart}       \\ \hline
\textbf{Human observer} \cite{seg-human+ASM+voting}             &        2048$\times$2048             & No                    &   -                 & -             & 0.946            & -             & 0.887            \\ \hline
\textbf{InvertedNet} \cite{seg-InvertedNet}               & 256$\times$256             & No                    & 3-fold   CV           & 0.974         & 0.950            & 0.937         & 0.882            \\ \hline
\textbf{u-net} by \cite{seg-u-net-brazil}     & 256$\times$256             & No                    & 5-fold    CV            & 0.976         & 0.962            &       -        &         -         \\ \hline
\textbf{u-net} by \cite{seg-u-net-scia}    & 256$\times$256             & No                    & 5-fold     CV        &       -        & 0.959            &       -        & 0.899            \\ \hline
\textbf{u-net} by us & 512$\times$512          & Yes                   & 5-fold      CV *          &       0.981        &       0.964           &      0.965         &        0.933          \\ \hline
\textbf{MTdG } (proposed)       & 512$\times$512             & Yes                   & 5-fold       CV  *      &          \textcolor{blue}{0.989}     &      \textcolor{blue}{0.978}          &     \textcolor{blue}{0.977}        &       \textcolor{blue}{0.956}          \\ \hline
\textbf{u-net} by \cite{seg-u-net-islam2018towards} &  512$\times$512          & Yes                   &   train/test split (80\%/20\%)**          &       0.986        &       -           &      -         &        -          \\ \hline
\textbf{SegNet} by \cite{seg-u-net-brazil}      & 256$\times$256             & No                    & 5-fold   CV           & 0.979         & 0.955            & 0.944         & 0.896            \\ \hline
\textbf{SCAN} \cite{seg-SCAN}         & 400$\times$400             & No                    &   train/test split (209/38)              & 0.973         & 0.947            & 0.927         & 0.866            \\ \hline
\textbf{FCN} by \cite{seg-u-net-brazil}     & 256$\times$256             & No                    & 5-fold CV             & 0.974         & 0.950            & 0.942         & 0.892            \\ \hline
\textbf{MTdG } (proposed)       & 256$\times$256             & No                   & 5-fold      CV        &          0.974     &      0.962          &     0.934        &       0.928          \\ \hline
\textbf{MTdG } (proposed)       & 256$\times$256             & No                   & 3-fold      CV        &          0.962     &      0.953          &     0.921        &       0.916          \\ \hline
\textbf{ASM tuned} \cite{seg-human+ASM+voting}     &       256$\times$256              &        No               &           2-fold    CV      &     -          & 0.927            &       -        & 0.814            \\ \hline
\textbf{Hybrid voting} \cite{seg-human+ASM+voting}     &        256$\times$256             &         No              &            2-fold CV         &      -         & 0.949            &      -         & 0.86             \\ \hline
\textbf{Seghers et. al} \cite{seg-Seghers}             &       256$\times$256              &          No            &            train/test split (50/44)        &        -       & 0.951            &        -       &       -           \\ \hline
\textbf{Ibragimov et. al} \cite{seg-Ibragimov}    &        -             &           No            &           -          &      -         & 0.953            &       -        &         -         \\ \hline
\end{tabular}
\begin{tablenotes}
      \footnotesize
      \item *: p-values from significance test are reported in Table \ref{tab:segmentation}.
      \item **: JSRT dataset is not used for this work.
\end{tablenotes}
\end{threeparttable}
\end{center}
\endminipage\hfill
\end{table*}

To the best of our knowledge, no multitask framework has been found in the literature to benchmark the proposed multitask network for our tasks. A comparison between \textit{u-net} (implemented by us) and \textit{pix2pix MTdG} along with the p-values of student's t-test is provided in Table \ref{tab:segmentation} which clearly shows that the \textit{pix2pix MTdG} method yields better results with a significant difference ($p < 0.001$). However, to contrast these results with other methods in the literature, due to the variations in utilizing different folding schemes and image sizes, a fair comparison of the results is not a straightforward process. 

The results of different state-of-the-art algorithms on the JSRT dataset are summarized in Table \ref{tab:seg-comp}. The settings reported from each method are also provided in Table \ref{tab:seg-comp} with `-' to mean that the value is not reported. For a fair comparison, the proposed method has also been tested using $256\times256$ image size, 3-fold cross-validation and without any augmentation. As presented in the Table, while the scores are really close to each other, the best-achieved results are the ones provided by \textit{pix2pix MTdG} in $512\times512$ image resolution. In all these other settings, the \textit{pix2pix MTdG} performance is still reasonable and is comparable to the performance of other techniques.  

\subsection{Task 2: Bone Suppression}
The second task of the \textit{pix2pix MTdG} network is bone suppression. The results of this task are evaluated via the structural similarity index (SSIM) metric for similarity estimation, and the root mean squared error (RMSE) metric to measure the difference between predicted and actual values \cite{ssim}. The RMSE measure between two X and Y images is expressed by Equation \eqref{eq:RMSE}, where $N$ in the total number of pixels in image and $i$ is the pixel index. 
SSIM is a reference-based quality assessment metric, which compares the local patterns of pixel intensities between the reference and output images. The maximum value of 1 implies that the two  images are structurally similar, while a zero value indicates that there is no structural similarity between them. Usually, the SSIM index is calculated via windowing on the images with 8x8 window size and 1 pixel striding. At the end, the mean of the computed values (M-SSIM) would be reported. The M-SSIM measure between two images $X$ and $Y$ and the default SSIM measure between two windows $W^X_i$ and $W^Y_i$ are as defined by Equations \eqref{eq:MSSIM} and \eqref{eq:SSIM} where $\mu_{W^X_i}$, $\mu_{W^Y_i}$, $\sigma_{W^Y_i}^2$ , $\sigma_{W^X_i}^2$, and $\sigma_{W^X_iW^Y_i}$ show the average of $W^X_i$, the average of $W^Y_i$, the variance of $W^X_i$, the variance of $W^Y_i$ and the Covariance of $W^X_i$ and $W^Y_i$, respectively. The default settings of $c_1=(0.01 \ L)^2$ and $c_2=(0.03 \ L)^2$ are considered with $L$ being the dynamic range of the pixel-values (i.e. 255 in our experiment).  

\begin{gather}
RMSE(X,Y)=\sqrt{\frac{1}{N} \sum_{i=1}^N (x_i-y_i)^2} 
\label{eq:RMSE}
\end{gather}

\begin{gather}
M{\text -}SSIM= \frac{1}{N} \sum_{i=1}^N SSIM(W^X_i,W^Y_i) 
\label{eq:MSSIM}
\\
SSIM(W^X_i,W^Y_i)= \hspace{120pt} \nonumber \\ \hspace{50pt} \frac{(2\mu_{W^X_i}\mu_{W^Y_i}+c_1)(2\sigma_{W^X_iW^Y_i}+c_2)}{(\mu_{W^X_i}^2+\mu_{W^Y_i}^2+c_1)(\sigma_{W^X_i}^2+\sigma_{W^Y_i}^2+c_2)}  
\label{eq:SSIM}
\end{gather}

\begin{figure}[]
\centering
\captionsetup{justification=centering}
  \includegraphics[width=0.9 \linewidth,trim={0 0 0 0cm},clip]{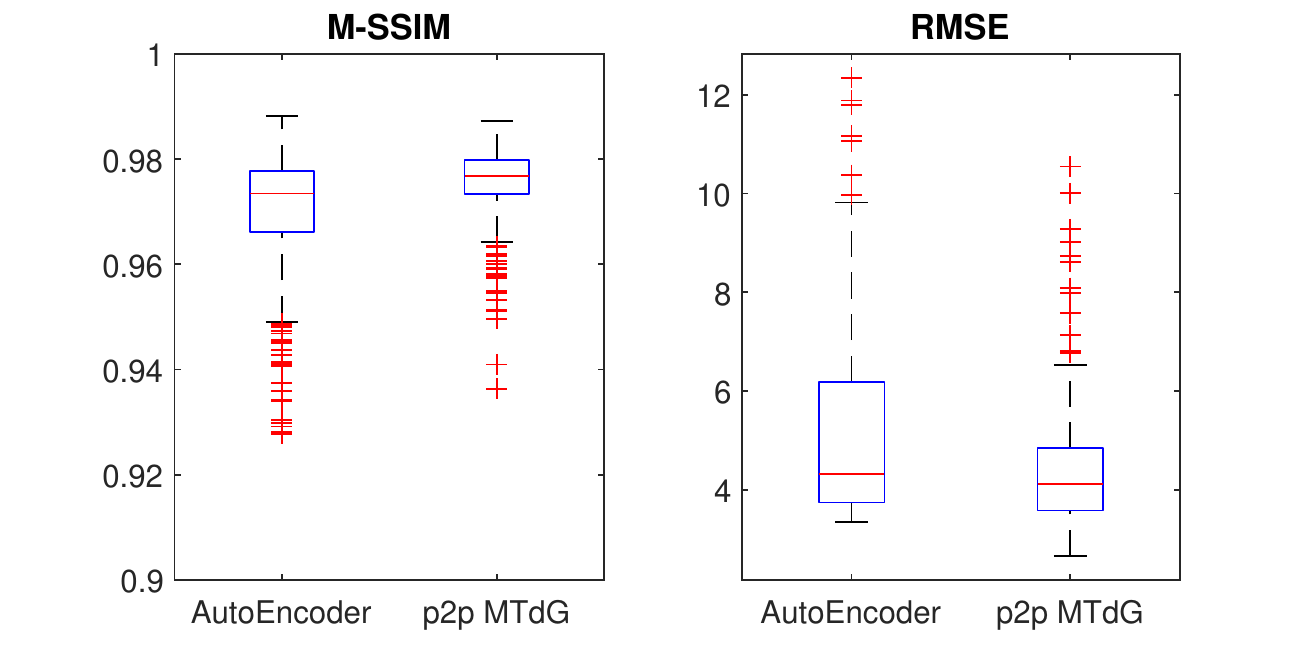}
  \caption{Bone suppression task showing boxplots of the results. \\ Left) M-SSIM similarity score. Right) RMSE difference.}
  \label{fig:SSIM}
\end{figure}

\begin{figure}[]
\centering
\captionsetup{justification=centering}
\minipage{0.5\textwidth}
  \includegraphics[width=\linewidth]{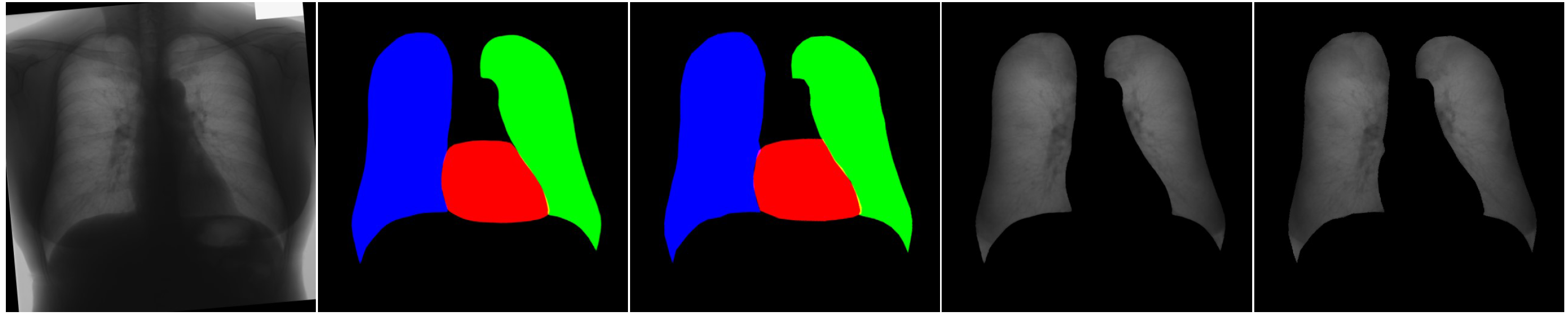}
\endminipage\vfill
\minipage{0.5\textwidth}
  \includegraphics[width=\linewidth]{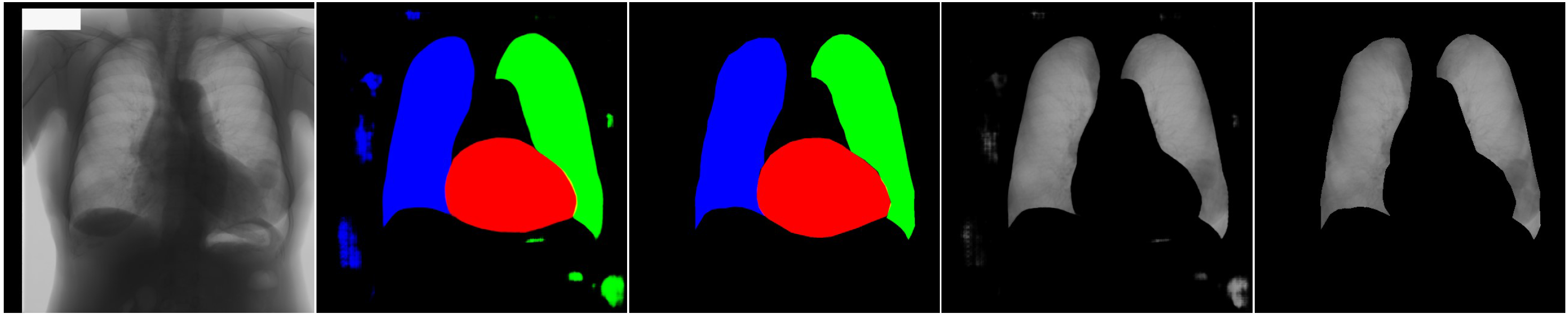}
\endminipage\vfill
  \caption{Results by proposed method \textit{pix2pix MTdG} regarding to the best (top) and worst RMSEs (bottom). \\
  \footnotesize{Columns left to right: input image, segmentation result, segmentation target, bone suppression result and bone suppression target. \\ 
  Top) Best RMSE: SSIM: 0.99, RMSE : 2.66. \\
  Bottom) Worst RMSE: SSIM: 0.94, RMSE : 10.55.} }
  \label{fig:image-rmse}
\end{figure}

\begin{table}[!t]
\centering
\caption{Results of bone suppression task via different methods.}
\label{tab:Bone}
\begin{tabular}{|c|c|c|c|}
\hline
              &  \textbf{AutoEncoder} \cite{bonesup-new-1} & \textbf{pix2pix MTdG}  & \textbf{p-value} \\ \hline
\textbf{MSSIM} &   0.970 $\pm$  0.011   &      \textcolor{blue}{0.976 $\pm$  0.006}    &        8.6 e-24              \\ \hline
\textbf{RMSE} &   5.096 $\pm$  1.812   &      \textcolor{blue}{4.297 $\pm$  1.046}     &            5.3 e-45           \\ \hline
\end{tabular}
\end{table}

\begin{table*}[!h]
\centering
\caption{Summary of the properties of different methods.}
\label{tab:summary}
\begin{tabular}{|c|c|c|c|c|c|c|c|}
\hline
                        \textbf{Task}     &                          & \textbf{pix2pix ST} & \textbf{pix2pix STdG} &  \textbf{U-net} & \textbf{AutoEncoder} & \textbf{pix2pix MT} & \textbf{pix2pix MTdG} \\ \hline
\multirow{2}{*}{\textbf{Segmentation}} & \textbf{Average Dice}       &  0.977 $\pm$ 0.008 &          0.984 + 0.008          & 0.976 $\pm$ 0.007  &         -              &  0.978 $\pm$ 0.008  &   \textcolor{blue}{0.985 $\pm$ 0.007}   \\ \cline{2-8} 
                                        & \textbf{Average FNR}       &   0.018 $\pm$ 0.008    &          0.014 + 0.009        &   0.017 $\pm$ 0.006    &      -             &  0.018 $\pm$ 0.008 &  \textcolor{blue}{0.012 $\pm$ 0.007}   \\ \hline
\multirow{2}{*}{\textbf{Rib Suppression}} & \textbf{Average MSSIM}  &   0.969 $\pm$ 0.009 &           0.974 $\pm$ 0.007       &       -            &  0.970 $\pm$  0.011  &  0.968 $\pm$ 0.007  &  \textcolor{blue}{0.976 $\pm$ 0.006}   \\ \cline{2-8} 
                                         & \textbf{Average RMSE}    &   5.296 $\pm$ 1.688   &       4.697 $\pm$ 1.587         &     -              &  5.096 $\pm$  1.812   &   5.382 $\pm$ 1.470   &  \textcolor{blue}{4.297 $\pm$ 1.046}    \\ \hline
\textbf{}                                & \textbf{No. Parameters}  &   \multicolumn{2}{c|}{2 $\times$ 57,190,084}            &  31,084,008        &         64,400             &  \multicolumn{2}{c|}{57,199,303}            \\ \hline
\textbf{}                                & \textbf{Minimum Epochs}  &   \multicolumn{2}{c|}{250 }                             &   48               &          150        &   \multicolumn{2}{c|}{300        }           \\ \hline
\textbf{}                                & \textbf{Training time}   &  \multicolumn{2}{c|}{786 = 2 $\times$ 393 min  }        &    248 min         &          852 min           &  \multicolumn{2}{c|}{662     min   }       \\ \hline
\end{tabular}
\end{table*}

While there is much literature about bone suppression in CXR images and even with the availability of proprietary software, there is no shared dataset available nor is there any open source codes or models that are shared with the research community. There is just one competitor study for bone suppression with shared source code in which the authors used an interesting \textit{AutoEncoder} network architecture \cite{bonesup-new-1}. 
The box plots of the achieved M-SSIMs and RMSEs for the \textit{pix2pix MTdG} and \textit{AutoEncoder} are shown in Figure \ref{fig:SSIM} and Table \ref{tab:Bone}, proving the good performance of the proposed \textit{pix2pix MTdG}. The best and worst results with regards to the RMSE measurement are shown in Figure \ref{fig:image-rmse}.

\subsection{Discussion}

As discussed earlier, the proposed \textit{pix2pix MTdG} method provides promising results in accomplishing both segmentation and bone suppression tasks simultaneously. In this subsection, other characteristics of the proposed method are discussed and a summary of these results is provided in Table \ref{tab:summary}.

\begin{figure*}[]
\captionsetup{justification=centering}
\minipage{0.33\textwidth}
  \includegraphics[width=\linewidth]{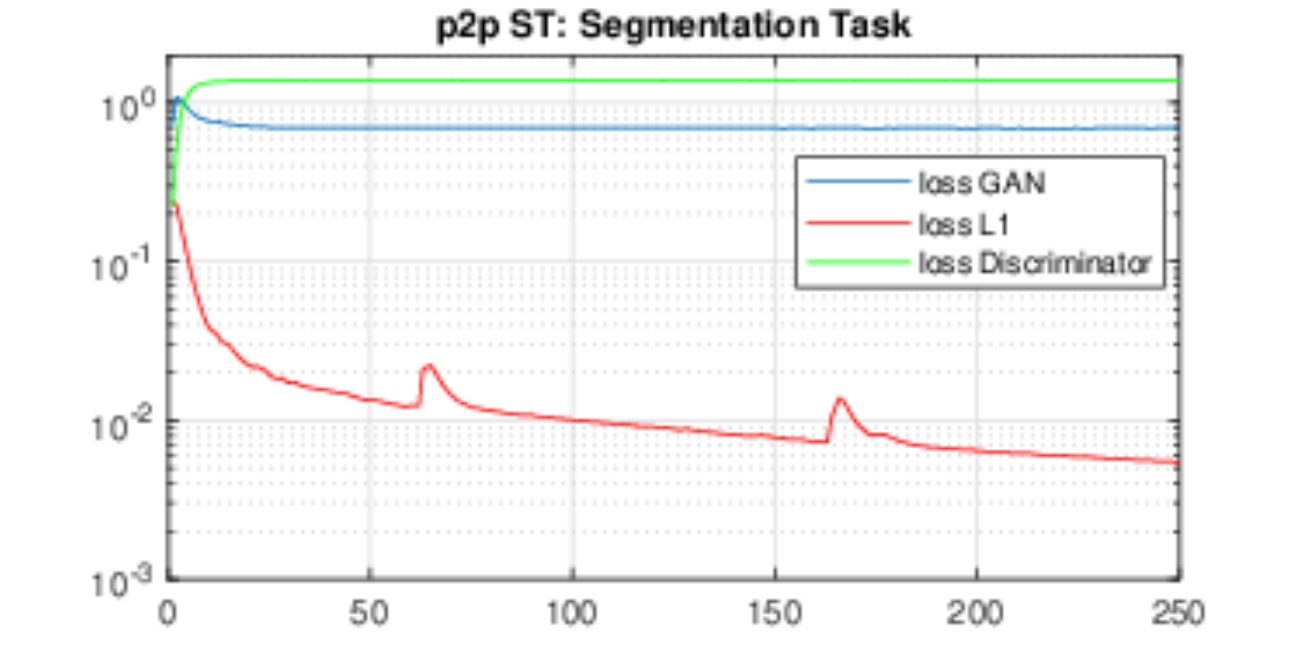}
\endminipage\hfill
\minipage{0.33\textwidth}
  \includegraphics[width=\linewidth]{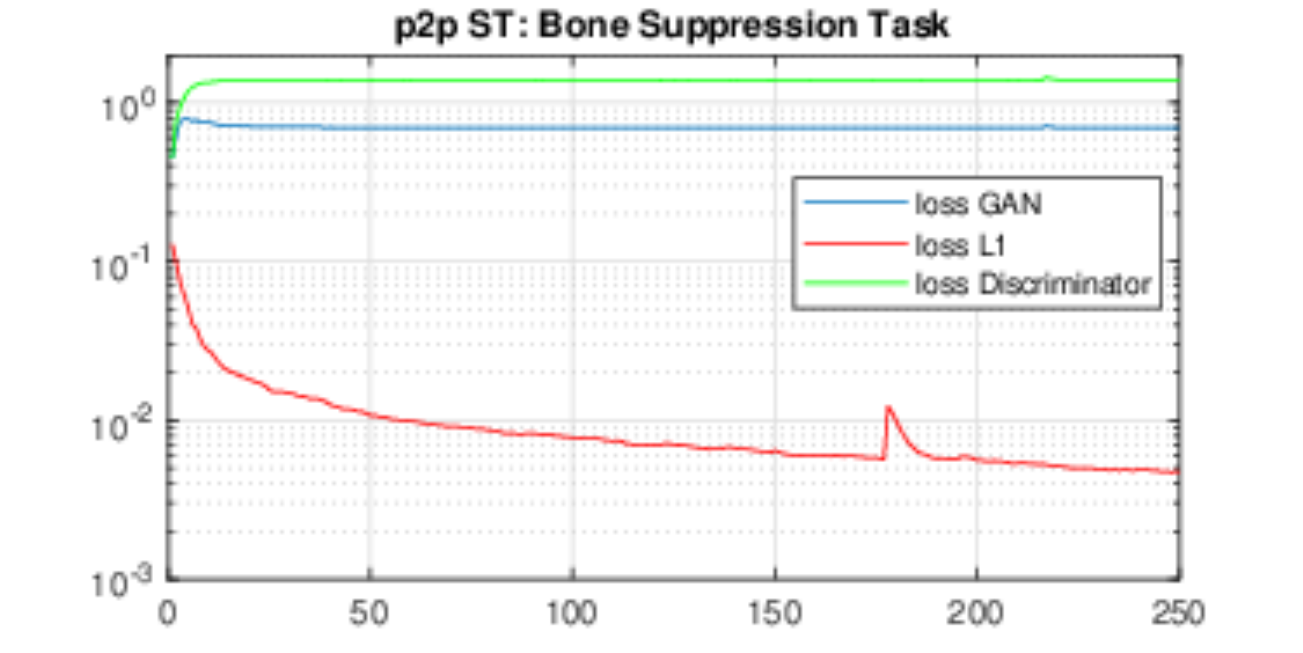}
\endminipage\hfill
\minipage{0.33\textwidth}
  \includegraphics[width=\linewidth]{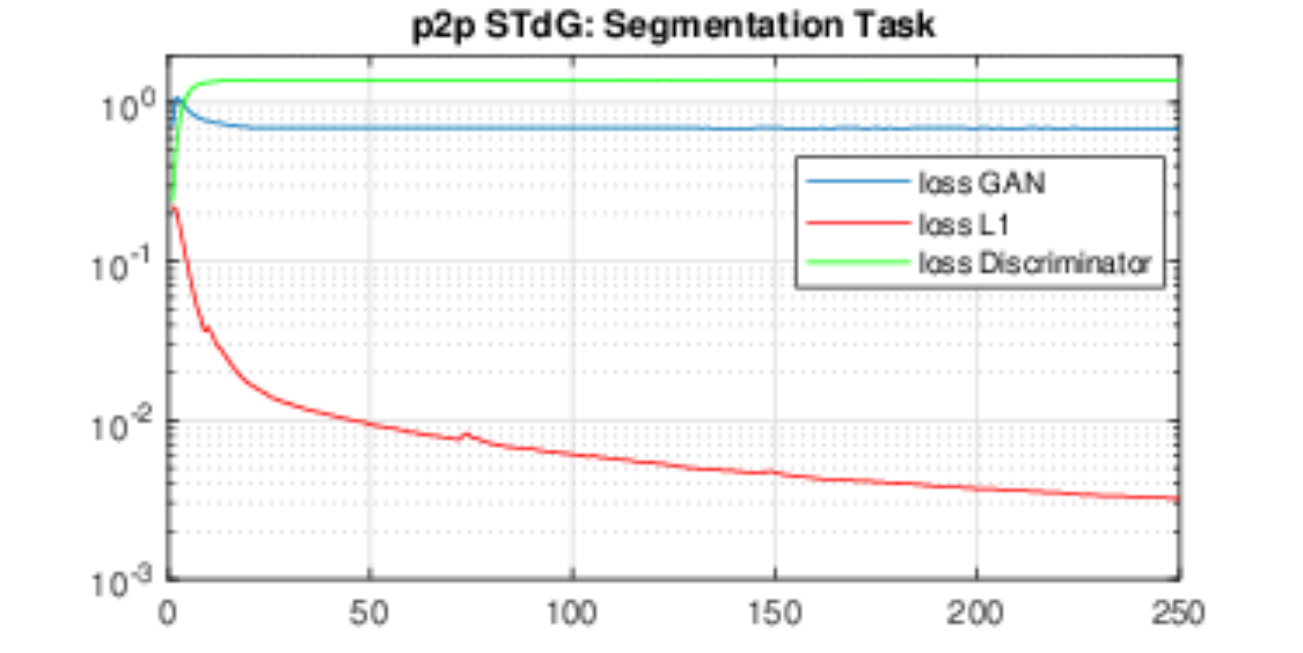}
\endminipage\hfill
\vfill
\minipage{0.33\textwidth}
  \includegraphics[width=\linewidth]{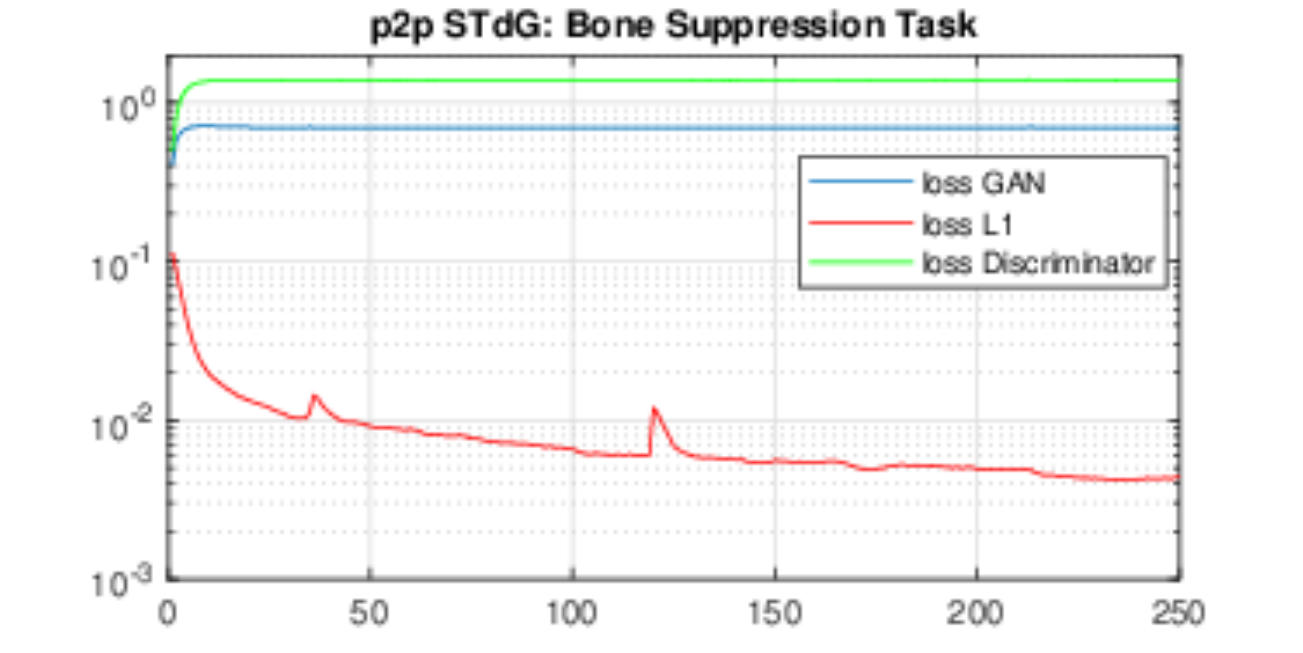}
\endminipage\hfill
\minipage{0.33\textwidth}
  \includegraphics[width=\linewidth]{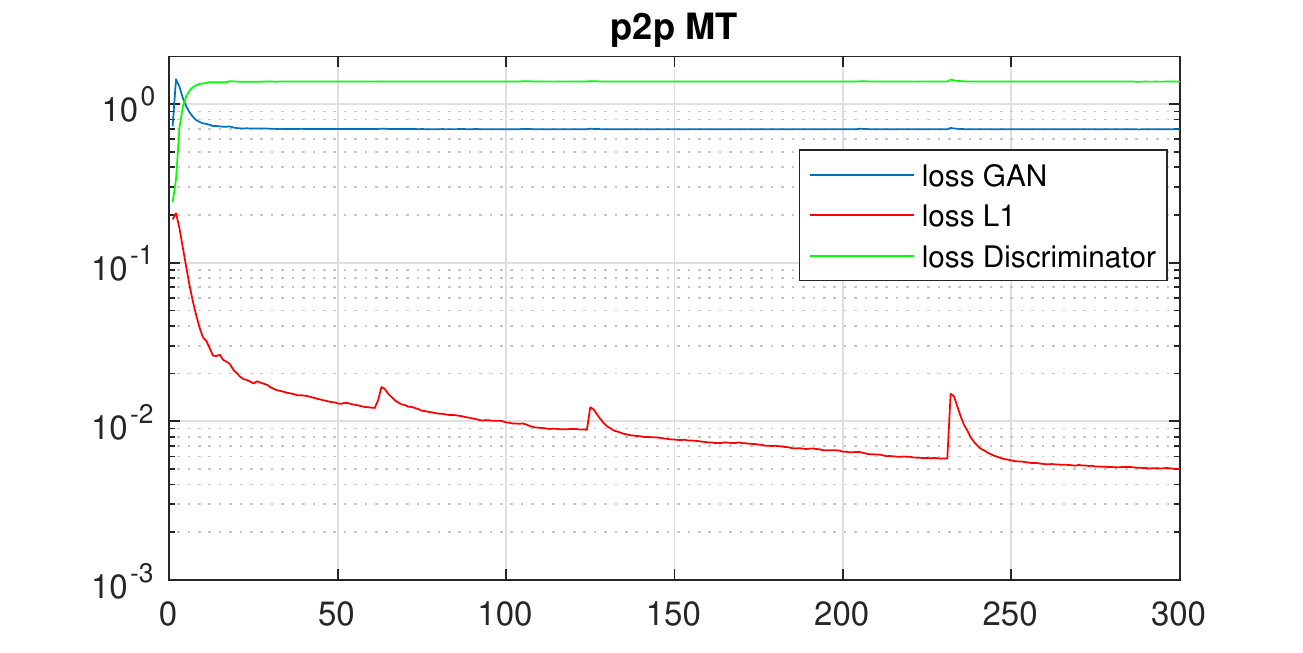}
\endminipage\hfill
\minipage{0.33\textwidth}
  \includegraphics[width=\linewidth]{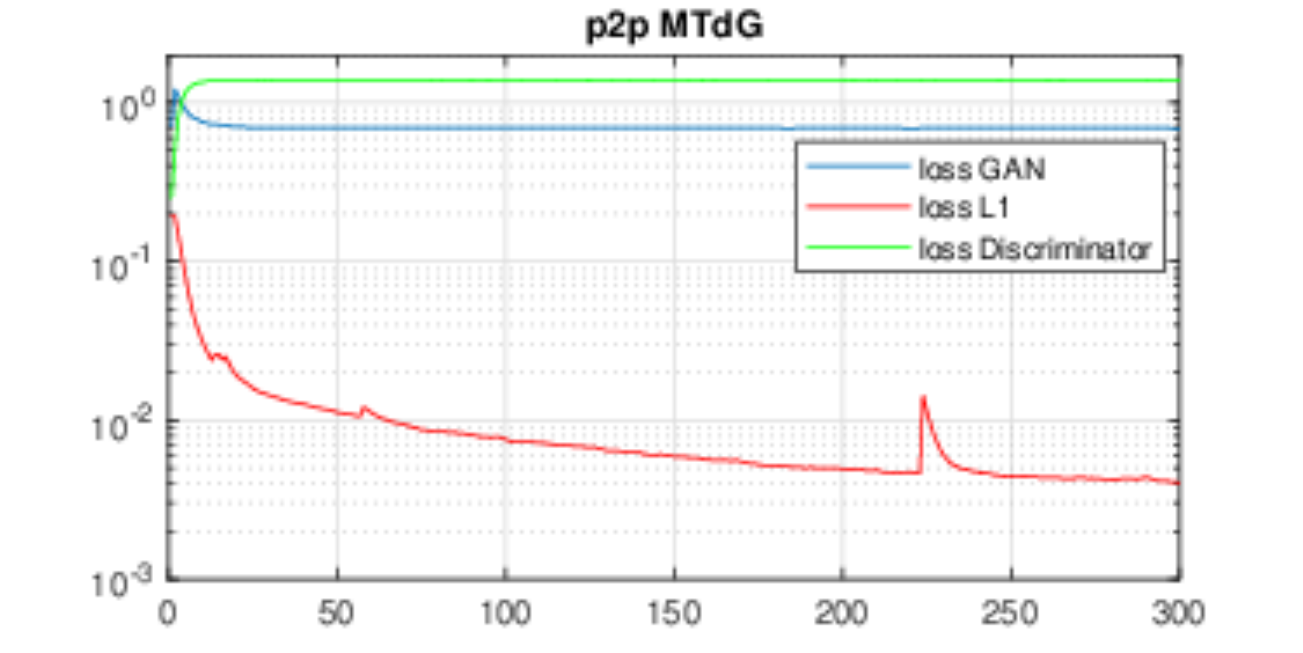}
\endminipage\hfill
  \caption{The loss curves of the different schemes for a training session as a function of epochs. \\ 
  \footnotesize{p2p: \textit{pix2pix}, ST: single task, STdG: single task with dilation in generator, MT: multitask, MTdG: Multitask with dilation in generator.}   }
  \label{fig:loss}
\end{figure*}

\subsubsection{Performance Analysis}
The performance of the proposed method is assessed here with respect to the results obtained and the network parameters that were considered. The following remarks can be made: \\
\textbf{a}) As shown in Tables \ref{tab:segmentation} and \ref{tab:summary}, the proposed multitask pix2pix with dilation (\textit{pix2pix MTdG}) achieves the best results for both organs segmentation and bone suppression tasks. \\
\textbf{b}) Using multitask pix2pix without dilation was not as effective for improving the results. \\
\textbf{c}) Another advantage of \textit{pix2pix MTdG} is in the number of required parameters of the network, which is an intrinsic requirement for the multitask pix2pix scheme. The trainable parameters of pix2pix, u-net, and multitask pix2pix are $57,190,084$; $31,084,008$ and $57,199,303$, respectively. Note that the u-net architecture can only perform the segmentation task. To employ two separate pix2pix networks for the two tasks, a large number of training parameters $2 \times 57,190,084$ would be needed. In other words, the performance of \textit{multitask pix2pix} is reasonable in the number of parameters used while it maintains comparable good results to the state-of-the-art techniques. \\
\textbf{d}) Nonetheless, the\textit{ multitask pix2pix }framework has one drawback, which is the number of required iteration/epochs for the training phase. In the experiments conducted in this study, \textit{pix2pix MTdG} required almost $300$ epochs while u-net and single task \textit{pix2pix} converged in only 50 and 250 epochs, respectively, meaning that the multitask framework requires more training time in comparison with other methods. It is worth noting that the above concern is just for training and since a pre-trained network is used to generate the output in the testing phase, both methods perform similarly fast requiring only 1.2 seconds.

\begin{figure*}[]
\captionsetup{justification=centering}
\minipage{0.33\textwidth}
  \includegraphics[width=\linewidth]{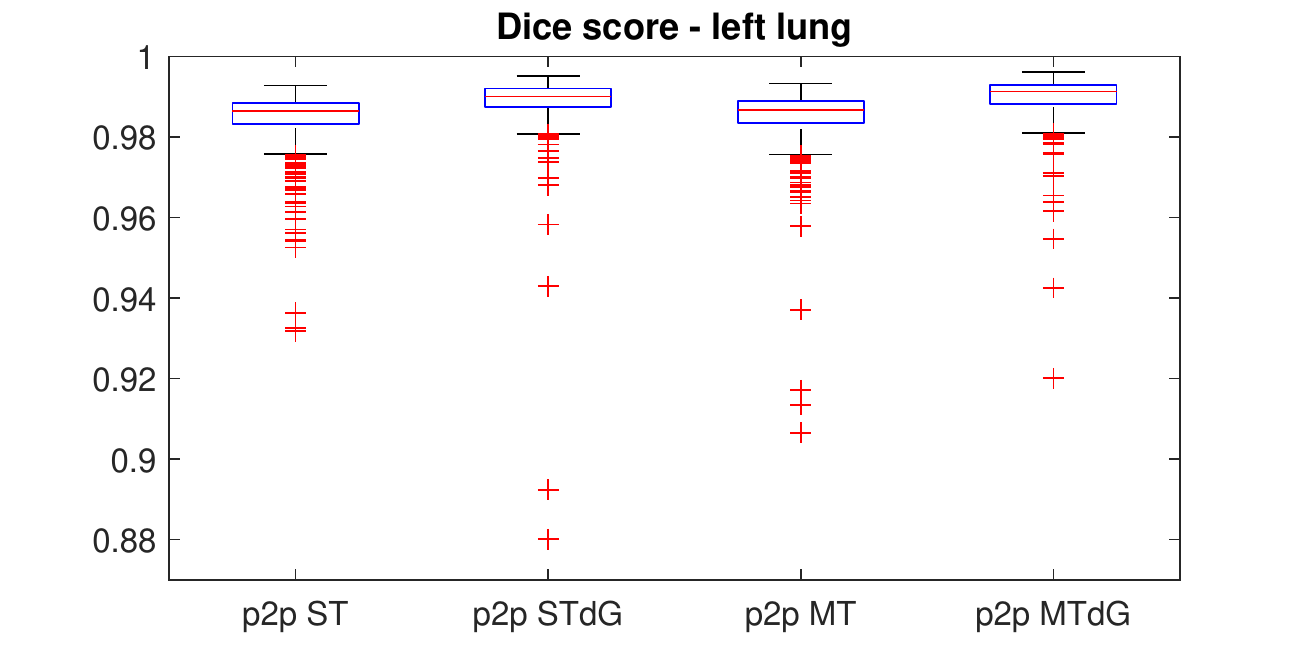}
\endminipage\hfill
\minipage{0.33\textwidth}
  \includegraphics[width=\linewidth]{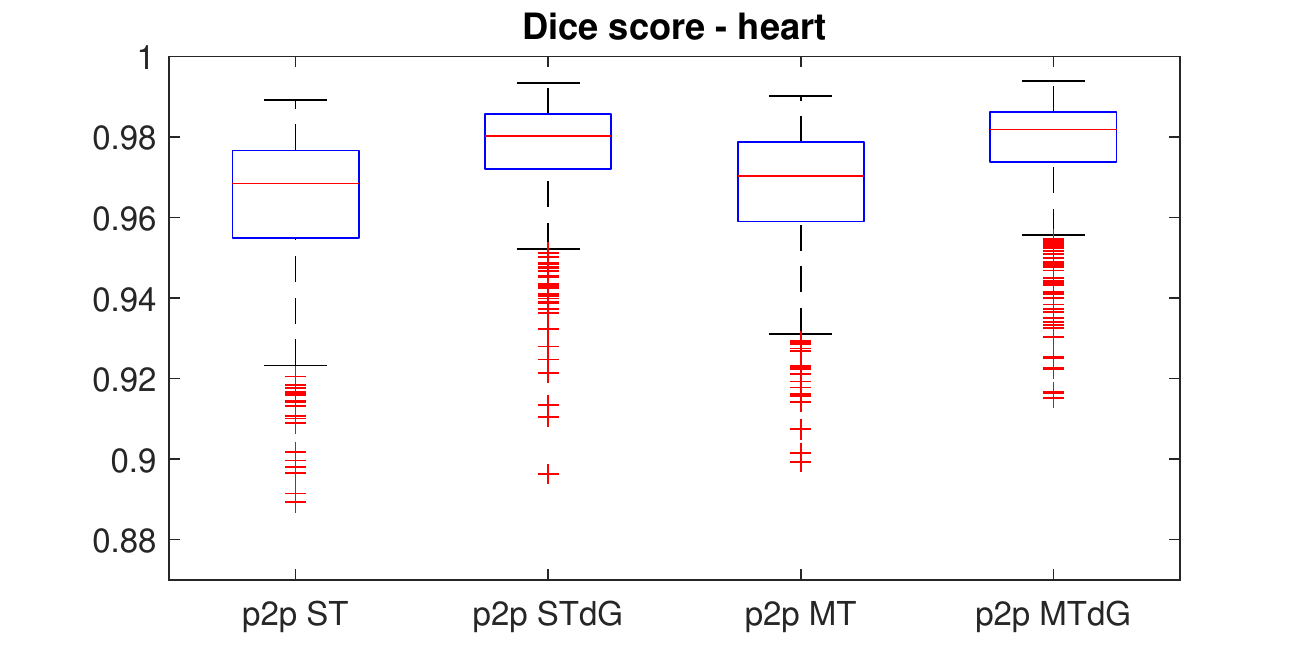}
\endminipage\hfill
\minipage{0.33\textwidth}
  \includegraphics[width=\linewidth]{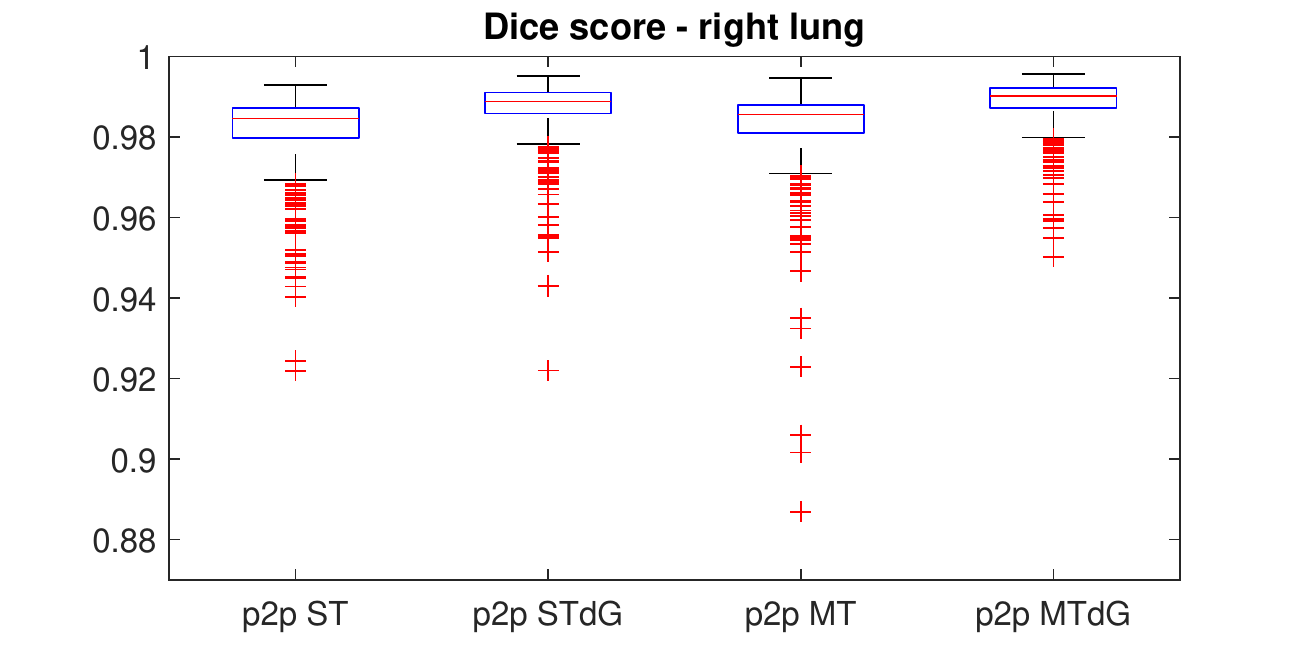}
\endminipage\hfill
\vfill
\minipage{0.33\textwidth}
  \includegraphics[width=\linewidth]{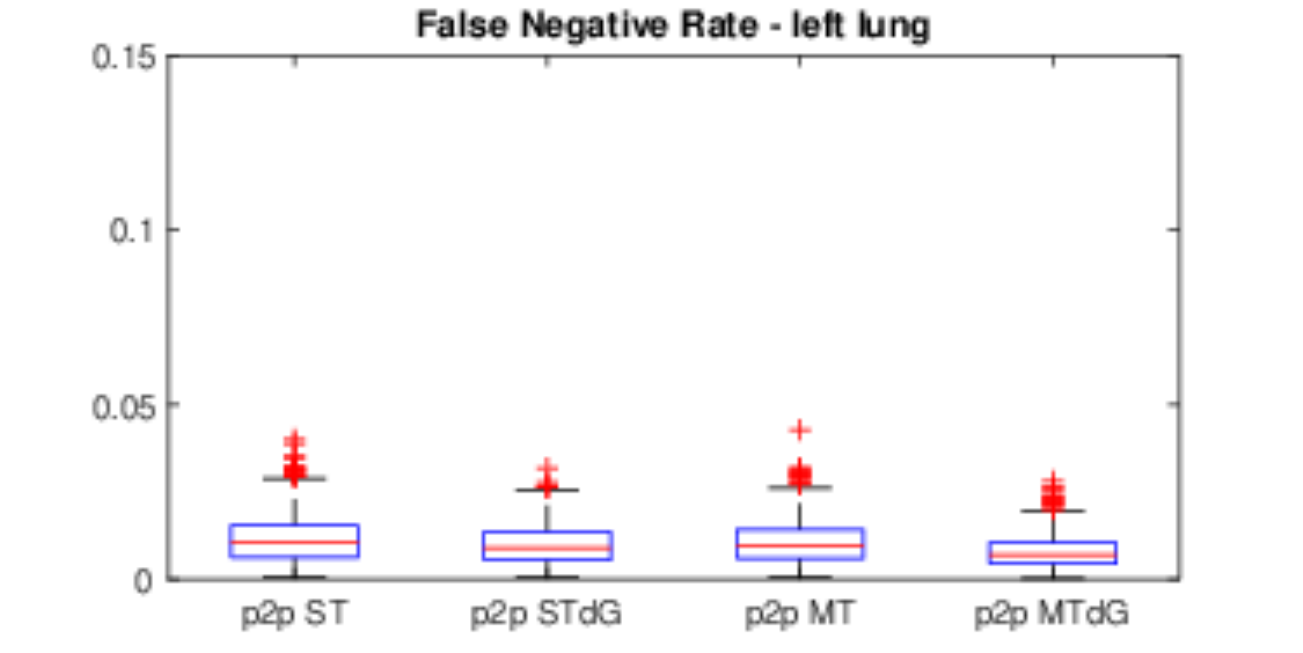}
\endminipage\hfill
\minipage{0.33\textwidth}
  \includegraphics[width=\linewidth]{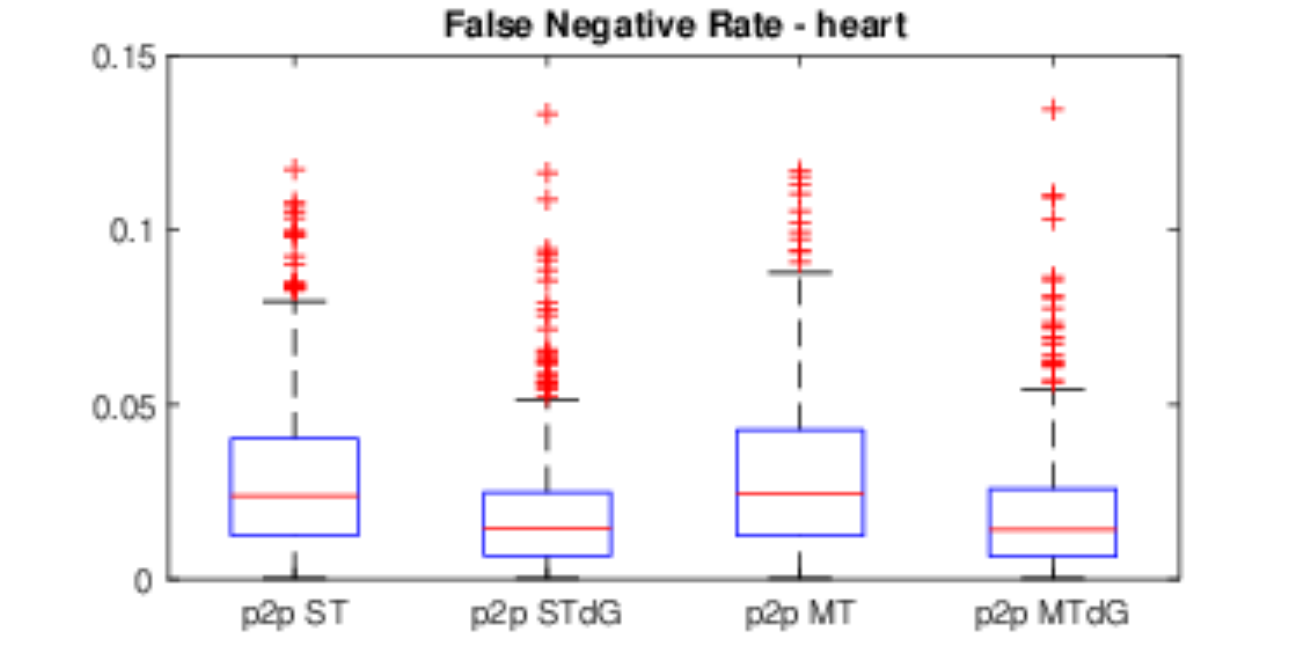}
\endminipage\hfill
\minipage{0.33\textwidth}
  \includegraphics[width=\linewidth]{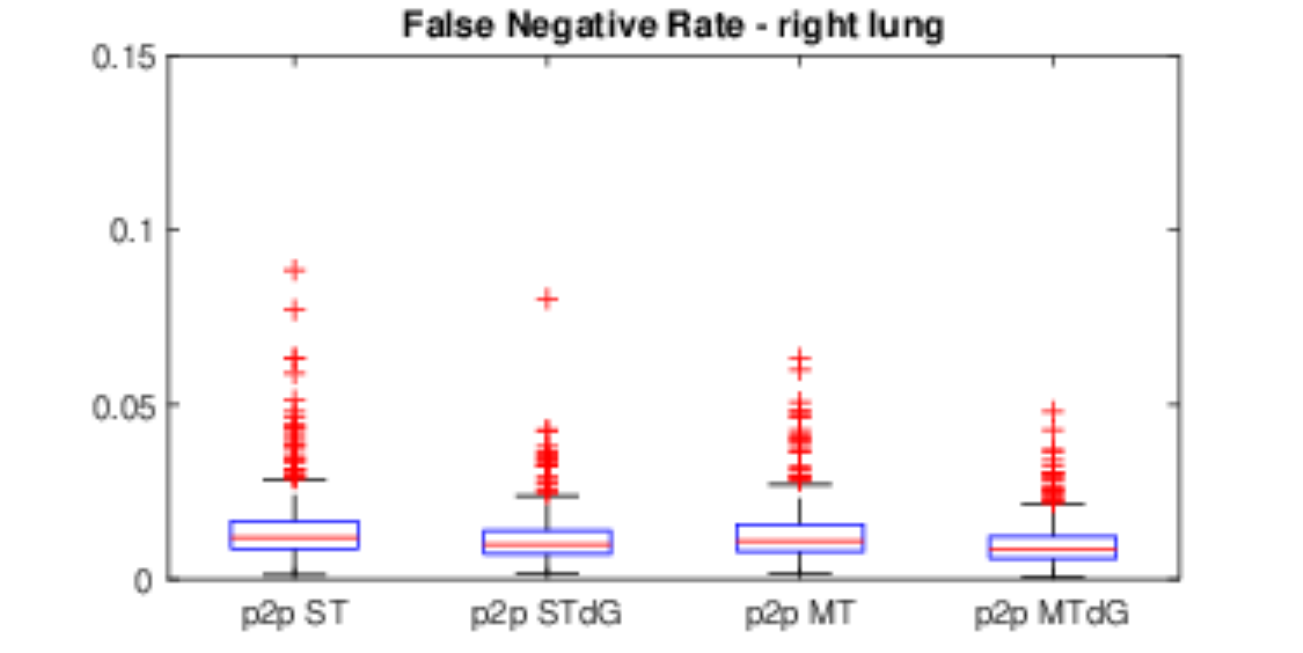}
\endminipage\hfill
  \caption{Segmentation task evaluation showing box-plots for different schemes. \\ (Left) left lung, (Middle) heart, (Right) right lung. \\ 
  \footnotesize{(\textit{pix2pix ST}, \textit{pix2pix STdG}, \textit{pix2pix MT} and \textit{pix2pix MTdG})}
  }
  \label{fig:ablation-T1}
\end{figure*}

\begin{figure}[]
\centering
\captionsetup{justification=centering}
\minipage{0.35\textwidth}
  \includegraphics[width=\linewidth]{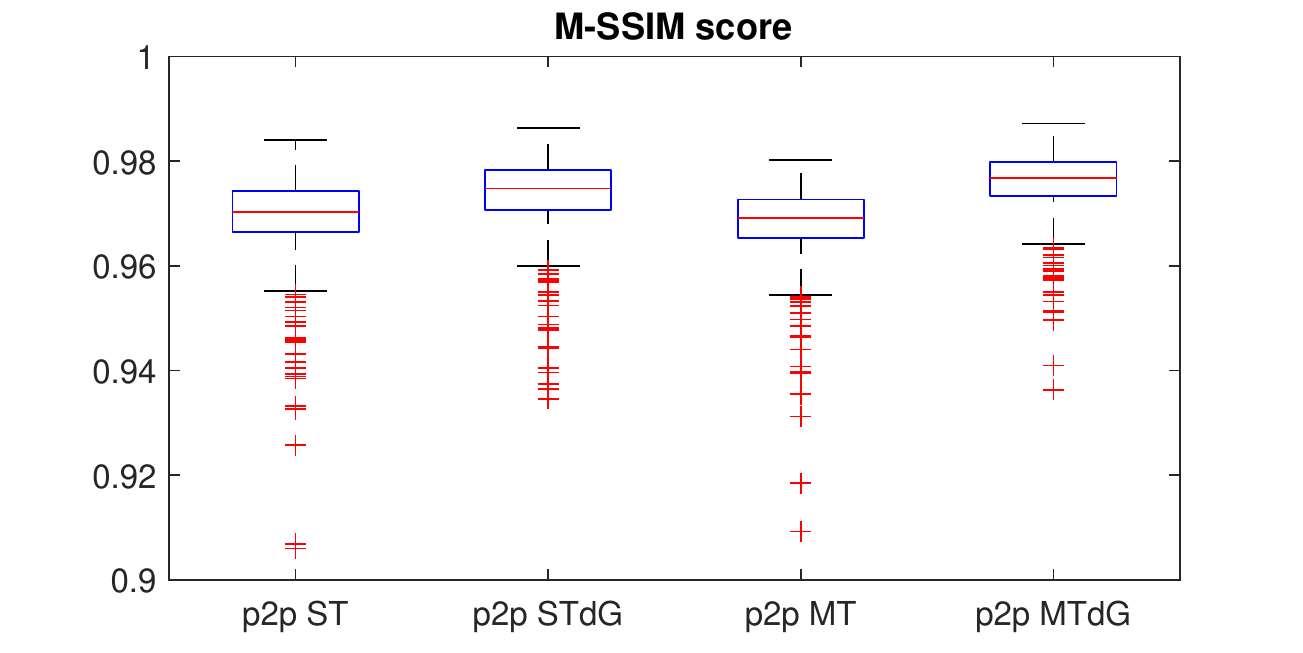}
\endminipage\vfill
\minipage{0.35\textwidth}
  \includegraphics[width=\linewidth]{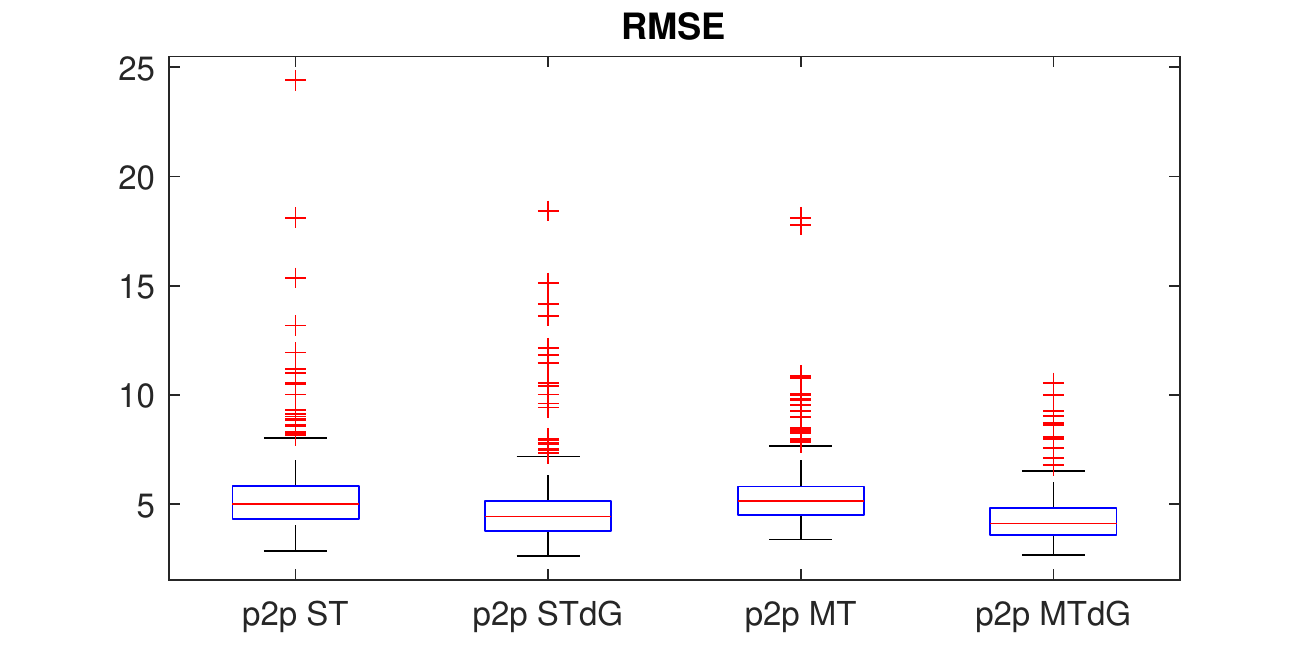}
\endminipage
  \caption{Bone suppression task evaluation showing the box-plots for the different schemes. \\ \footnotesize{(\textit{pix2pix ST}, \textit{pix2pix STdG}, \textit{pix2pix MT} and \textit{pix2pix MTdG})} }
  \label{fig:ablation-T2}
\end{figure}

\begin{figure}[]
\centering
\captionsetup{justification=centering}
\minipage{0.5\textwidth}
  \includegraphics[width=\linewidth]{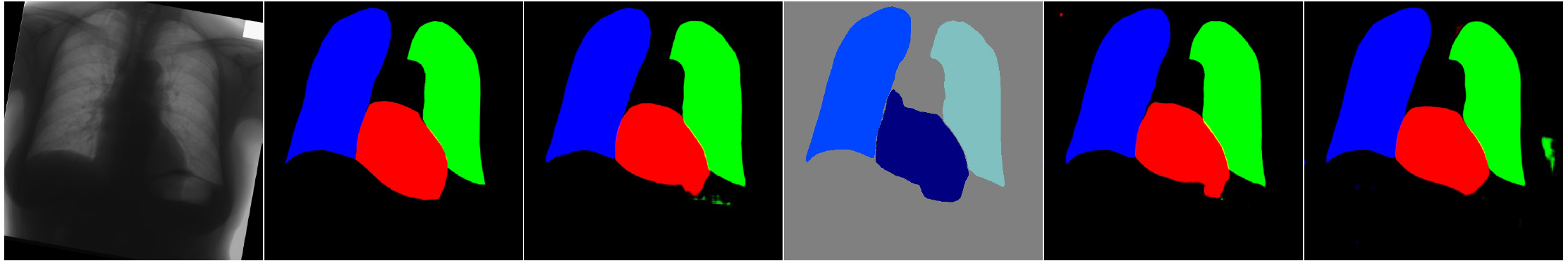}
\endminipage\vfill
\minipage{0.5\textwidth}
  \includegraphics[width=\linewidth]{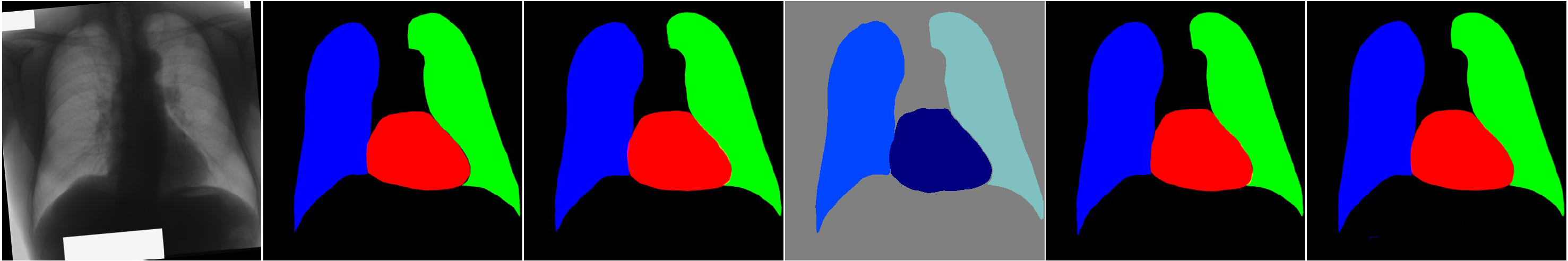}
\endminipage\vfill
\minipage{0.5\textwidth}
  \includegraphics[width=\linewidth]{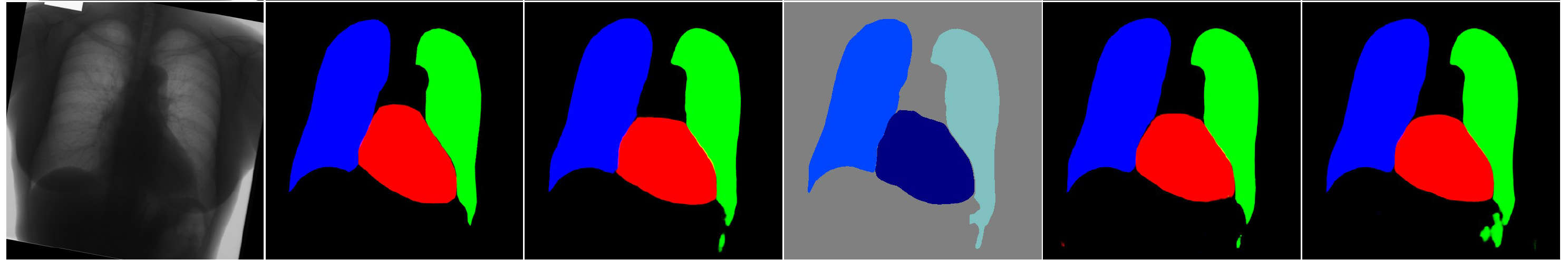}
\endminipage\vfill
\minipage{0.5\textwidth}
  \includegraphics[width=\linewidth]{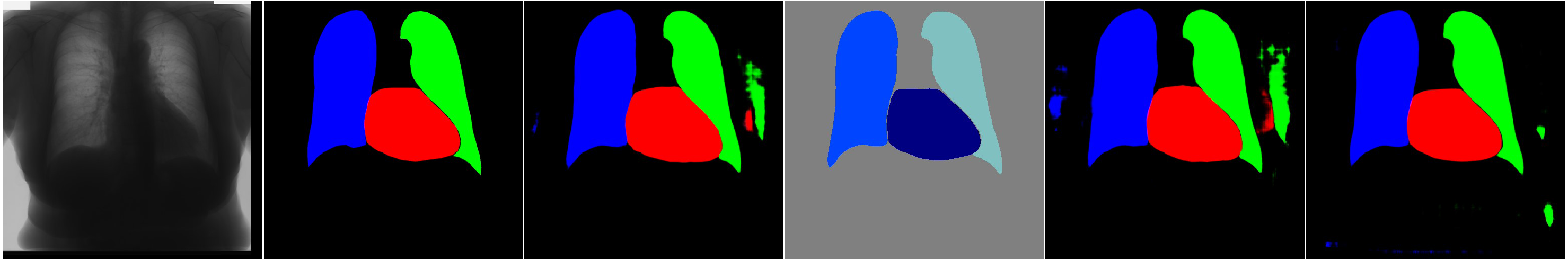}
\endminipage\vfill
  \caption{Segmentation results of different schemes for different subjects. Top to bottom are subjects and left to right are input, target, \textit{pix2pix ST}, \textit{u-net}, \textit{pix2pix MT} and \textit{pix2pix MTdG}.}
  \label{fig:comparison}
\end{figure}

\subsubsection{Ablation Study}

In order to investigate the effectiveness of dilation and multitasking, the ablation study is considered. For this reason, six processing schemes are addressed: two \textit{single task pix2pix} (\textit{p2p ST}) for two tasks, two \textit{single task pix2pix with dilation} (\textit{p2p STdG}), \textit{multi task pix2pix} (\textit{p2p MT}) and \textit{multitask pix2pix with dilation} (\textit{p2p MTdG}). Figure \ref{fig:loss} shows the loss curves for the training session of the schemes with respect to epochs. The loss GAN, loss L1 and loss discriminator expressed as $\mathbb{E} [-log(F +\epsilon )]$, $\mathbb{E} [\mid Y - \hat{Y} \mid_1 ] $ and $\mathbb{E} [-( \ log(R + \epsilon) + log(1-F + \epsilon ) \ )]$, respectively, were included in the loss functions \eqref{eq:OurLoss-G} and \eqref{eq:OurLoss-D}. The training is stopped when the L1 loss reaches and stabilizes at almost $0.005$. As shown in Fig. \ref{fig:loss}, all of the schemes are able to converge and make the generator the declared winner. Furthermore, while the multitask schemes need more epochs for reaching the desired L1 loss value, there is no significant difference between the achieved final loss values.

\begin{figure*}[!hb]
\centering
\captionsetup{justification=centering}
\includegraphics[width=\linewidth, trim={0 23cm 0cm 0},clip]{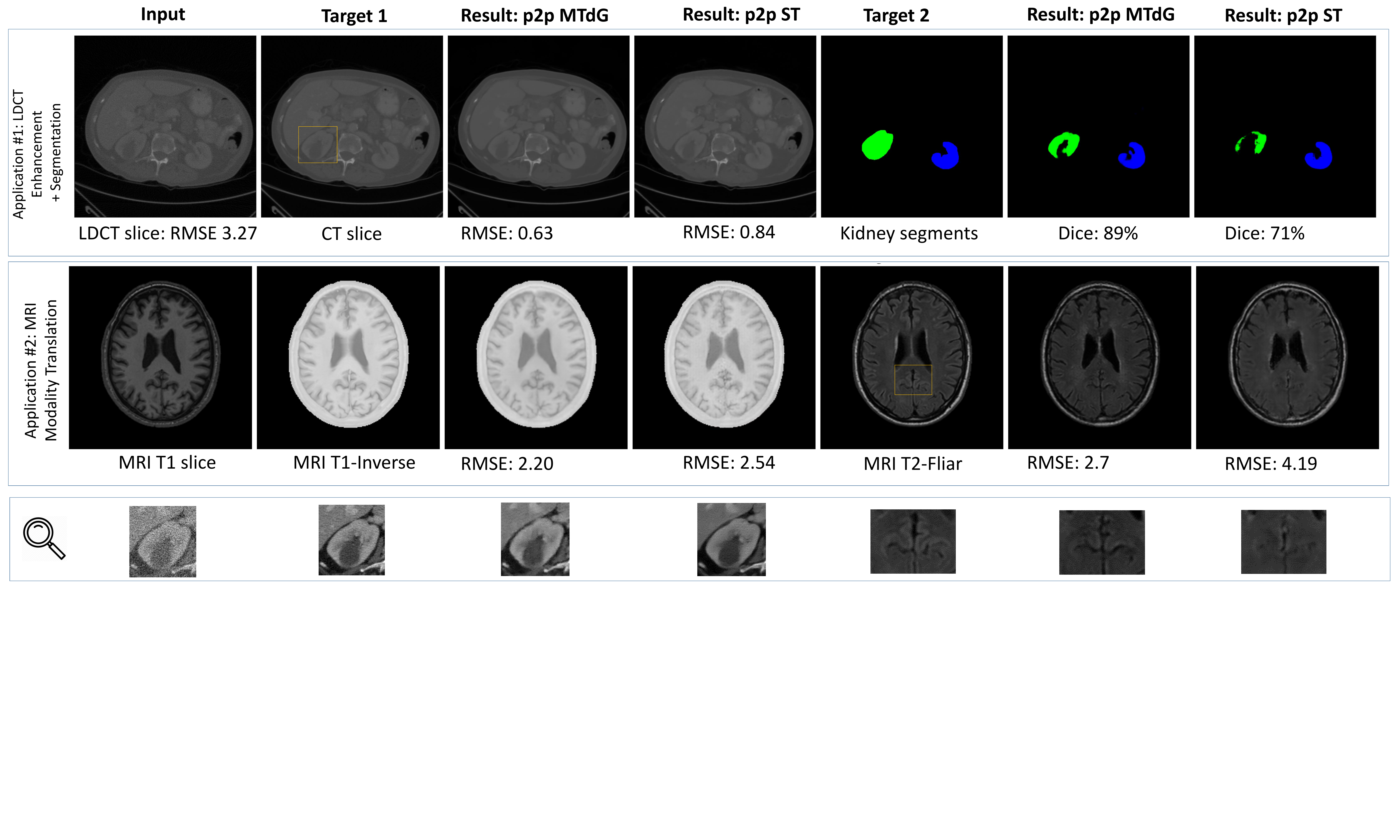} 
  \caption{Preliminary results of two more applications of \textit{multitask pix2pix} for joint tasks. \\
  Top row) Low dose CT: image enhancement and segmentation. \\
  Mid row) MRI Neuroimage translation: T1 to T1-inverse and T2-flair. \\
  Bottom row) Magnification of the selected region (area inside yellow rectangle). \\
  \footnotesize{p2p: \textit{pix2pix}, ST: single task, MTdG: Multitask with dilation in generator.}  }
  \label{fig:generalization}
\end{figure*}

In contrast to the almost similar final training loss values of the different schemes, as shown in Table \ref{tab:summary}, there are differences in the test sessions and in the outputs of the schemes' models. The best results are achieved by \textit{pix2pix MTdG} while the multitasking without dilation was not helpful. This could be inferred as a hyper-parameter selection criterion, as the dilation makes the receptive fields bigger to become more suitable to the nature of the CXR input images and the organs' shapes and sizes. On the other hand, when dilation is exploited in the network, multitasking (\textit{p2p MTdG}) assumes the benefits and generates better results than those obtained from a single task with dilation (\textit{p2p STdG}); this outcome is achieved despite the fact that the multitask network has half of the weights in comparison to two single-task networks. Figures \ref{fig:ablation-T1} and \ref{fig:ablation-T2} demonstrate the results of segmentation and bone suppression tasks for different schemes as can be observed from the box-plots of Dice score, false negative rates, RMSE and M-SSIM score in concurrence to the results shown in table \ref{tab:summary}, especially when comparing for outliers, percentile edges and median line.

\subsubsection{Qualitative Analysis}

Figure \ref{fig:comparison} presents the segmentation results of the different methods for various subjects. From left to right, the images shown are the input image, the target image, single task \textit{pix2pix} output, \textit{u-net} output, \textit{pix2pix MT} output, and \textit{pix2pix MTdG} output. In retrospect, the following assessments can be made:\\
\textbf{a}) As presented in the first row, the \textit{pix2pix MTdG} framework delivered the best outcomes in comparison with the other methods. \\
\textbf{b}) However, the result of all the other methods in the second row is almost the same. \\
\textbf{c}) The third row shows that the \textit{multitask pix2pix} without dilation achieved better results in contrast to the other methods. This could be related to the fact that this dilation process may not be as effective in some specific cases. We intend to combine the layers with and without dilation as future work to see if the accuracy could be improved further. \\
\textbf{d}) Moreover, as it is illustrated in the fourth row of Figure \ref{fig:comparison}, the \textit{pix2pix} based methods suffer from false-positive segments such as isolated islands. Although this is the case for the u-net method as well,  u-net demonstrates better results in this case. This could be associated with the fact that the loss function of u-net is addressing the segmentation constraints while the loss function of \textit{pix2pix} is constructed to perform the pixel-wise comparison. This drawback could simply be removed by employing post-processing techniques such as connected components and considering the island area, which is not the aim of this paper at this juncture, but could be exploited in future work. Another approach to deal with this drawback is to implement segmentation related loss functions such as dice score to \textit{pix2pix} loss function. While this technique was discussed by some authors in the literature, our investigations did not prove its efficiency in delivering any improvement to our case.

\subsubsection{Generalization of Method and Future Work}
\label{subsec:generalization}
The size of the input CXR images considered in this study are 512x512. This resolution chosen for the proposed method is  also exploited in the Stanford's \textit{CheXNext} network \cite{general-Chest-Andrew} with promising results.  Nonetheless, manufacturers of X-ray radiography continue to improve its resolution and hence 512x512 could be limiting in obtaining similar optimal results sought with a higher resolution. Since more pixels mean more revealing information that could enhance the prospects for more accurate diagnosis, it is thus reasonable to extend the proposed network for dealing with higher resolution images such as 2048x2048. It should be noted that since the generator of the proposed method is a holistic network (not locally and patch or block-based), it is expected to yield more promising results with higher resolution images, but at the expense of more convolutional layers and with a higher demand for more variables to contend with and hence more taxing computational requirements. 

With these contending challenges in mind, and in order to show the potential benefits of the \textit{image-to-images} translation and \textit{multitask pix2pix} with respect to the generalization aspect to other domains of application, even under the 512x512 resolution, this section provides preliminary results on two more experiments that were conducted. The first experiment involves low dose CT image enhancement and segmentation of kidneys. The second is a neuroimaging translation for cross-modality generation of  T2-flair and T1-inverse from the T1 input image. 

Close attention is given in the literature to LDCT imaging  because of the use of a lower dose of radiation and its wider availability for being affordable with faster scanning time making it suitable for screening, diagnosis and follow up visits \cite{ldct-nature-lung}. While segmentation is an intrinsic problem in imaging, likewise in LDCT, the image enhancement aspect that could lead to better image quality is also a state-of-the-art issue  which is being addressed in the literature \cite{ldct-nature-rec}. 

In the second experiment, while \textit{Cross-modality generation} can serve as an auxiliary method in clinical diagnosis \cite{n2n_general}, it also has great potential for  multimodal registration  \cite{n2n_registration, n2n_arxiv,n2n_arxiv_seg_treg}, segmentation \cite{n2n_registration, n2n_arxiv_seg_treg, n2n_segmentation}, super-resolution \cite{n2n_superresolution} and structural information improvement \cite{n2n_enhancement} (e.g. MRI 3T to 7T). For this second experiment, an MRI T1 volume of the brain is translated into an MRI T1-inverse and a T2-flair, in slice by slice fashion. 


Figure \ref{fig:generalization} shows the qualitative and quantitative results (as seen in one slice) of single task and \textit{multitask pix2pix} using leave one subject out (LOSO) as a test evaluation scheme. The network is trained using all of the 2D axial slices from the training subjects and is tested on all of the 2D axial slices of the test subjects not seen in the training phase. For these studies, the hyper-parameters and networks are the same as the aforementioned for CXR analysis and are not optimized for these applications. For both applications, the multitask method outperforms the single task methods while having only half of the network weights. It is worth mentioning that there is a significant difference in segmentation task for LDCT application and T2-flair task for the neuroimaging application. The magnification area is shown to emphasize the differences. These results clearly show the potential benefits of using \textit{image-to-images} translation to other domains of application involving different imaging modalities even when the proposed model as used for these additional applications are implemented on the model solely based on CXR images.   


%% file: SecConclusion.tex
\section{Conclusion}
\label{sec-conclusion}

Of all the many existing medical imaging modalities, X-ray imaging remains the most widely used modality as it is the most cost effective and one of the easiest to administer. Chest X-ray remains an essential imaging modality for the diagnosis and follow up treatment of many diseases affecting the lungs, heart and bone structure within the chest area. In this study, a new deep learning based \textit{image-to-images} approach was proposed that simultaneously suppresses the bones that hinder the visibility and scrutiny of organs and nodules and segments the organs within the chest area. Essentially, and for the first time, the architectural design of this deep learning-based model exploits in the most effective way the interplay of parameters in between the two tasks to optimize the outcome for both tasks at once. In order to perform these two essential tasks of bone suppression and organs segmentation,  the well-established pix2pix network is extended to generate two output images simultaneously (an image with bones suppressed, and a second image showing the segmented organs), yielding the new \textit{image-to-imag\underline{es}} with an automated end-to-end framework instead of the traditional \textit{image-to-image} approach that deals with each task separately. The proposed method was trained via an end-to-end process and is evaluated by cross validation and significance testing with several standard metrics, resulting in highly accurate results for both tasks. Through two additional empirical evaluations involving low-dose CT images and neuroimaging, the proposed architectural design of the model is shown to be amenable for generalization to other domains of application, although developed around CXR imaging.  
In summary, the contributions of this work can be summarized as follow:
\begin{itemize}
    \item This work is the first to try to extend the application of \textit{image-to-image} network to \textit{image-to-images} network, while optimizing the use of parameters and securing computational efficiency.
    \item The network is improved through the inclusion of dilated convolutions in some specific layers, which are shown to improve the accuracy of the results significantly. 
    \item The \textit{image-to-images} network is used to accomplish simultaneously the two common and most needed tasks of bone suppression and organ segmentation in CXR images while optimizing the outcomes for both.
    \item All of the developed code is shared publicly online for validation purposes and for use by the research community interested in the automated diagnosis and in treatment follow up using chest X rays.
\end{itemize}